\documentclass[a4paper, usenatbib]{mnras}
\usepackage[utf8]{inputenc}
\usepackage{graphicx}
\usepackage{natbib}
\usepackage{multicol}
\usepackage{amsmath}	
\usepackage{amssymb}
\usepackage{mathrsfs}	
\usepackage{bm}

\usepackage[T1]{fontenc}

\def\apj{ApJ}
\def\apjl{ApJL}
\def\apss{Ap\&SS}
\def\aap{A\&A}

\def\mnras{MNRAS}

\def\prd{PhRvD}
\def\prl{PhRvL}
\def\sovast{SvA}
\def\sovastl{SvAL}
\def\astl{AstL}
\def\aplett{ApL}

\def\beq#1{\begin{equation}\label{#1}}
\def\eeq{\end{equation}}
\def\beqa#1{\begin{eqnarray}\label{#1}}
\def\eeqa{\end{eqnarray}}

\def\eqn#1{(\ref{#1})}

\title[Radiation-dominated polar emitting region of an accreting \mbox{X-ray} pulsar -- I]{Radiation-dominated polar emitting region of an accreting \mbox{X-ray} pulsar -- I. Polarization- and spectrum-dependent structure, and the emergent continuum}
\author[M. Gornostaev] {M. I. Gornostaev\thanks{E-mail: mgornost@gmail.com}\\
Sternberg Astronomical Institute, Lomonosov Moscow State University, Universitetskij pr. 13, Moscow 119234, Russia\\
Faculty of Physics, Lomonosov Moscow State University, Leninskie Gory 1-2, Moscow 119991, Russia
}

\begin{document}

\date{Accepted XXX. Received YYY; in original form ZZZ}
\pagerange{\pageref{firstpage}--\pageref{lastpage}} \pubyear{}

\maketitle

\label{firstpage}

\begin{abstract}

 The  radiation-dominated  polar  emitting region of an accreting \mbox{X-ray} pulsar is simulated numerically  in the framework of a three-dimensional (geometrically two-dimensional)  model. The radiative transfer within the  emitting region and the structure of the latter are calculated with the use of the self-consistent algorithm developed earlier. The magnetic scattering cross sections dependent on the photon energy and polarization have been  incorporated. Second-order bulk Comptonization over entire emitting region, induced Compton scattering, the switching of the polarization modes, free-free processes, the cyclotron emission because of electron-proton collisions, and a realistic shape of the accretion channel have been taken into account. The case of a dipole magnetic field is considered.  It is shown that the induced Compton effect can play a notable role in establishing the electron temperature in the post-shock  zone.
  Within the model shock wave, a higher electron temperature  is  achieved  than  in the post-shock zone by means of the bulk-heating mechanism.
 The photons gaining the energy in the shock wave and above it due to bulk motion effects and the thermal Doppler effect are responsible for the formation of  high-energy regions in the emergent continuum of the polarization modes.

\end{abstract}


\section{Introduction} \label{sec:intro}

An accreting \mbox{X-ray} pulsar \citep{1971ApJ...167L..67G} is an included in a binary system, magnetized neutron star  on to which a plasma inflows  from a companion star (e.g.~\citealt{1972A&A....21....1P,1973ApJ...179..585D,1973A&A....22..421B}).  The accreting material is directed towards the neutron star along the magnetic field lines, increasing the kinetic energy,  the main fraction of which is eventually released as the energy of \mbox{X-ray} photons emerging from heated to  corresponding temperatures regions of the atmosphere of the neutron star  or those grown above the star. Usually, the areas of the emitting regions across the magnetic field are much smaller than the area of the neutron star surface, and  the magnetic and rotational axes of the neutron star are far enough apart to coincide. Hence, the rotation of the neutron star provides the possibility of the  formation of the pulses of the \mbox{X-ray} radiation that comes from those regions to a distant observer.

When the mass accretion rate is sufficient, above the neutron star surface, inside the accretion channel, the region containing  the slowed plasma flow appears having, on average, a notably larger transverse Thomson optical depth than the freely falling flow. Such structures are often referred to as the accretion columns.  Under certain circumstances, the radiation pressure in the structure is much greater  than the gas pressure. Photons diffused from the almost stagnate interiors  are then advected by the compressed heated flow to the neutron star surface (diffusing at the same time across the magnetic field), to a region near the side boundary of the channel, so that such a structure turns out to be  surrounded by a radiation-dominated shock wave  \citep{1973ApJ...179..585D, 1973NPhS..246....1D,  1950PhRv...80..692D, 1958PhFl....1...24M}.

It is well known that modelling the deceleration of the accretion flow in a radiation-dominated shock wave is an example of a problem, that has not yet been considered exhaustively in the framework of a unitary study. The 2D numerical solution of \cite{1973NPhS..246....1D} became the first quantitative theoretical  result showed the structure of the emitting region formed in sufficiently luminous  accretion-powered \mbox{X-ray} pulsars. A one-dimentional approximation of the initially 2D equations from this work was considered by \cite{1976MNRAS.175..395B}. \cite{1981MNRAS.194.1033B} derived in the Fokker-Planck approximation the radiative transfer equation \citep{1960ratr.book.....C,1978stat.book.....M}  for the compressed flow of a plasma.
This result is used in most of the papers on the spectral radiative transfer in the accretion columns listed in the next two paragraphs. In the papers by \cite{1981MNRAS.194.1041B} and \cite{1982SvAL....8..330Leng}, the solutions for the spectrum of radiation of a 1D radiation-dominated shock were studied.

\cite{1981A&A....93..255W} obtained a number of 2D numerical solutions for the steady-state structure of the column.  \cite{1985A&A...142..430K} developed a 2D analytical solution for the structure of the radiation-dominated column at luminosities at which the optically thick zone of the column has a relatively small height.   \cite*{1987ApJ...312..666A} described the processes taking place in a strongly magnetized plasma under the conditions of accretion columns.  The first computational steps to solve the problem in a self-consistent way were carried out in the following three works. 
\cite{1988ApJ...327..760R} considered  in the grey approximation a self-consistent 1D problem of the dynamics, electron temperature and radiative transfer inside the column.
\cite{1988A&A...205..215R} modelled the 2D  column structure and solved the 3D (geometrically 2D) radiative transfer equation for the mode-summed spectral intensity in an uncoupled manner, and the frequency-dependent diffusion coefficients were used. 
\cite{1989A&A...225..137R} obtained a 3D self-consistent solution describing the 2D  structure of the column and the total emergent spectrum, and the frequency-integrated energy equation was included in the computational procedure. Thermal Comptonization was neglected and  bulk motion effects of second order in the velocity were not taken into account.  Only  cases of relatively low accretion rates were considered, at which the radiation pressure effects become dynamically significant, so that typical obtained absolute values of the  divergence of the bulk velocity  turned out to be relatively small. The angular distributions of the intensity of radiation emergent from the sidewall of the accretion channel were obtained with the use of the method of last scattering  by \cite{1989A&A...223..246K}.   \cite{1989A&A...223..251M} studied the effects of the bulk motion on the radiation patterns from the side boundary of the flow formed at different frequencies solving a simplified coherent radiation transfer equation under the conditions of the moving material.

\cite*{1991ApJ...367..575B} calculated the spectra of the normal polarization modes from a preset radiation-dominated mound with taking into account the angular photon redistribution using the Feautrier method \citep*{1988ApJ...324.1001B,1990ApJ...349..262B}, and the processes in the shock wave and above were neglected.  The inclusion of the photon bubble oscillations  in the numerical calculations of the radiation-dominated accretion column was performed by \cite{1989ESASP.296...89K} (see also \citealt{1992ApJ...388..561A,1996ApJ...457L..85K}; \citealt*{1997ApJ...478..663H}). A steady-state 1D dynamical model of the cylindrical column was developed by \cite{1998ApJ...498..790B}. Analytical solutions for the radiative transfer equation  represented in the Fokker-Planck approximation for such a column were obtained  by \cite{2005ApJ...630..465B, 2007ApJ...654..435B} with making use of the velocity profiles allowing the separation of variables, but different from linear ones. Numerical solution of the spatially 1D radiative transfer equation with the use of the fenomenological power-law velocity profiles was carried out by \cite{2012A&A...538A..67F,2016A&A...591A..29F}. 
\cite{2015MNRAS.447.1847M} obtained an expression for the critical luminosity and some solutions describing the structure of the column. \cite*{2017ApJ...835..129W,2017ApJ...835..130W} self-consistently simulated the column taking into consideration the equations of state for the two-component plasma. The structure of the flow was supposed to be one-dimensional.  \cite*{2021MNRAS.501..564G} (hereafter referred to as G2021) solved numerically the problems of the 3D dynamical structure of the inclined accretion  columns, and (in a self-consistent way) the geometrically 2D problem of the structure of the column and its emergent spectrum, regardless of the straightforward observational applications (the gas pressure was neglected).  The MHD simulations of the 2D structure of the column were performed by \cite*{2021MNRAS.508..617Z, 2022MNRAS.515.4371Z}, and development of the photon bubble instability was confirmed. \cite*{2022ApJ...939...67B} took into account the upward widening of the accretion channel, analysed different velocity profiles (including those for the gas-dominated column), and added to their previous considerations some calculations of the cross sections in a strong magnetic field.

The structure of the accretion column should be interpreted as primarily dynamical one in  the description above. There are not many attempts not based on thermalization-related predictions using the Stefan-Boltzmann law to calculate the thermal properties of the column.
 Although establishing the temperature of electrons inside the radiation-dominated plasma in accordance with the local Compton equilibrium was mentioned repeatedly (e.g.~\citealt{1981MNRAS.194.1033B, 1982SvAL....8..330Leng, 2007ApJ...654..435B, 1970JETPL..11...35Z,1975SvA....18..691I}), the use of the specified approximation in calculations of the structure of the accretion column is quite rare (\citealt{1988ApJ...327..760R, 2017ApJ...835..129W, 2017ApJ...835..130W}; G2021). Often the temperature was set constant and parametrized (e.g.~\citealt{2007ApJ...654..435B,2022ApJ...939...67B}).

 It seems that a physically consistent modelling of the radiative transfer and accretion column structure taking into consideration  the entire diversity of non-stationary processes, inhomogeneities in the gas flow, the influence of the gas component, and including the calculation of cyclotron processes and  polarization of the radiation is a multidimensional problem which still does not  have  complete solution  \citep{1987ApJ...312..666A}.
Many years have passed since the estimates of \cite{1973ApJ...179..585D} appeared to be useful for thorough progress in understanding the nature of the observed phenomena. There are the observational features of the accretion-powered \mbox{X-ray} pulsars that need to be explained in the framework of theoretical models embracing as much properties of the real objects as possible. The formation of the spectrum within the polar emitting regions is related  with two interesting manifestations (among others): variations in the continuum and in the parameters of the cyclotron resonant scattering feature (CRSF) \citep{1978ApJ...219L.105T}. 
It is remarkable that the experimental evidences obtained suggest the existence of the relationship between the changes in the continuum and  CRSF \citep{2011A&A...532A.126K}.

The positive correlation of the CRSF energy with the \mbox{X-ray} luminosity $L_{\rm X}$ is often associated with the decrease (increase) in the height of the emitting region (being in the inhomogeneous magnetic field of an \mbox{X-ray} pulsar)  at increasing (decreasing)  $L_{\rm X}$ \citep{2007A&A...465L..25S, 2017MNRAS.466.2752R,2006MNRAS.368..690L}. Such a property of the emitting region is often explained \citep{2017MNRAS.466.2752R,2017A&A...601A.126V, 2019A&A...622A..61S} by the formation of the gas-dominated accretion column bounded from above by a collisionless shock wave (\citealt{1970SvA....13..566B, 1975ApJ...198..671S,1982ApJ...257..733L, 1950PhRv...80..692D}). \cite{2015MNRAS.454.2714M} proposed the mechanism for the positive correlation related to the Doppler shift of the CRSF during transitions between regimes of a relatively rapid flow and the radiation-dominated accretion column.
The negative correlation of the CRSF energy with $L_{\rm X}$ \citep{1990ApJ...365L..59M, 2006MNRAS.371...19T,2017MNRAS.466.2143D,2018A&A...610A..88V} is usually related with the emitting region being in the radiation-dominated regime. However, there is no complete explanation for the phenomenon, and the number of the observed sources demonstrating such a dependence is very small. Possible known interpretations include: variations in the characteristic magnetic field strength due to an increase (decrease) in the height of the emitting region with increasing (decreasing) $L_{\rm X}$; variations in the magnetic field strength within the accretion funnel due to the MHD instabilities \citep*{2013MNRAS.430.1976M}; a significant influence on the CRSF position of the intra-atmospheric magnetic field of the neutron star during  reflecting the direct radiation from the accretion funnel containing a column of variable height \citep{2013ApJ...777..115P}.

The variety of the accumulated observational data on the \mbox{X-ray} continuum of pulsars in a high-luminosity state has also not yet found a complete interpretation. A part of the observations  contain the spectra with a fallen power-law tail extending from relatively low energies~\mbox{$\lesssim 10$~keV}  \citep{1991ApJ...367..575B, 2007ApJ...654..435B,2017ApJ...835..130W,2022ApJ...939...67B}. For some observations, the spectra are characteristic having a high-energy thermal hump, which is sometimes added by a pronounced fallen quasi-power-law tail \citep{2017MNRAS.466.2143D,2022ApJ...932..106K,2024MNRAS.528.7320S}.

In this work I model the shape of the continuum from the radiation-dominated accretion column of an \mbox{X-ray} pulsar. It is well known that a strongly magnetized plasma is a birefringent medium where the polarized radiation exists in the form of two normal modes (\citealt{1970PhRvL..25.1061A}; \citealt*{1971PhRvD...3.2303C,1970pewp.book.....G}). The electric vector of one of them, called the extraordinary mode, oscillates mainly perpendicularly to the plane originated by the wavevector and magnetic field direction, and the electric vector of another mode, called the ordinary mode, oscillates mainly in the specified plane. Here, the limit of large Faraday depolarization is considered, which is the case when the modes are orthogonal to high accuracy \citep{1974JETP...38..903G, 1992hrfm.book.....M}.  The length-scales and magnetic field values (much less than \mbox{$10^{14}$~G}) within the problem  allow using this approximation without concern for the off-diagonal components of the intensity matrix.

The new solutions are numerically  obtained within a self-consistent  model of an axisymmetrical radiation-dominated accretion column developed in the 3D space (originated by two spatial variables and the photon energy). The calculated dynamical and thermal structures of the column are polarization- and spectrum-dependent. The radiative transfer equations are written in the Fokker-Planck approximation for the extraordinary and ordinary normal modes with the use of frequency-dependent magnetic scattering cross sections.  The results include the 2D distributions of the bulk velocity, the electron number density, the total radiation energy density, the electron temperature calculated in the local Compton equilibrium approximation, and related to those  distributions of the photon occupation number for both polarization modes. 
The spectral luminosity of the emergent radiation and, consequently, the polarization fraction are calculated from specified distributions of the occupation numbers.

 The basic equations,  parameters of the problem and numerical method are described in Section~\ref{sec:calc}. The results are presented in Section~\ref{sec:res}. The discussion is given in Section~\ref{sec:d}, and conclusions are contained in Section~\ref{sec:concl}.

\section{Model setup} \label{sec:calc}

 It will be useful in what follows to consider the flow of plasma accreting on to one of the magnetic poles of a neutron star. Let $\dot M$ be the mass accretion rate per magnetic pole characterising this flow.   It is supposed that the plasma is purely hydrogen and fully ionized.
Since the source of the energy of \mbox{X-rays} from the polar emitting region is the gravitational energy of the accreting matter, a luminosity of the region cannot exceed at a given $\dot M$ the value of
\beq{e:Laccr}
L\sim \frac{GM\dot M}{R},
\eeq  
where $M$ and $R$ are the mass and radius of the neutron star, correspondingly, and $G$ is the gravitational constant.  
A radiation-dominated mound-like structure forms in the  accretion funnel when $L$ exceeds the critical value \mbox{$L_{\rm cr}\sim 10^{36}$--$10^{37}~{\rm erg~s}^{-1}$} \citep{1976MNRAS.175..395B, 1981A&A....93..255W,2007ApJ...654..435B,2015MNRAS.447.1847M}.  I will consider in detail the mechanisms responsible for the conversation of the kinetic energy of the flow, and consequently for the formation of the  structure of the column and the \mbox{X-ray} continuum from the boundary of the funnel.

\subsection{Basic equations}

Since the shape of the accretion channel is determined by the geometry of the magnetic field, it is useful to introduce spherical polar coordinates \mbox{($r$, $\theta$, $\phi$)} appropriate to the special situation under consideration \citep{1981A&A....93..255W}.  The frame is inertial and related with the neutron star, 
and $r$ is counted from the centre of the star. At the star surface \mbox{$r=R$}. Above the surface $r$ is counted along the magnetic field lines. At the upper boundary  of the considered region of the funnel \mbox{$r=r_{\rm up}$}.   The symmetry about the magnetic axis corresponding to \mbox{$\theta=0$} is supposed (all quantities are independent 		of~$\phi$). Within the funnel, the magnetic field strength \mbox{$B\sim B_{\rm ns}r_*^{-l}$}, where $B_{\rm ns}$ is the strength at the neutron star surface, \mbox{$r_*=r/R$}, and $l$ is determined by the shape of the magnetic field lines. 
 Let $\theta_0$ be the angle defined by the side boundary of the filled funnel, and $\theta_1$ and $\theta_2$ be the angles defined by the inner and outer side boundaries of the hollow channel, correspondingly.

  The steady state of the accretion column is of interest. The momentum equation reads 
\citep{1973NPhS..246....1D}
 \beq{e:mom}
 n_{\rm e}m_{\rm p}({\bm v} \nabla)\bm v=-\frac{\nabla u}{3},
 \eeq 
 where $\bm v$ is the bulk velocity,  $u$ is the total mean (angle-averaged) radiation energy density, $n_{\rm e}$ is the electron number density, and $m_{\rm p}$ is the proton mass.   
 Equation \eqn{e:mom} is written under the assumption that dynamics of the flow is fully determined by the radiation, so that the right-hand side is the radiation pressure gradient accurate to the sign. The gravitational acceleration, the gas pressure, and the viscosity are neglected.   The mean  total radiation energy density is
  \beq{e:dens}
 u=\int\limits_0^\infty u_\epsilon {\rm d}\epsilon,
  \eeq 
  where $u_\epsilon={8\pi}\epsilon ^3 n/({c^3h^3})$ is the mean spectral radiation energy density, with  \mbox{$n=n_1+n_2$} is the  mode-summed mean photon occupation number, and $n_1$ and $n_2$ are the mean photon occupation numbers for the extraordinary (subscript~`$1$') and ordinary (subscript~`$2$') modes (other quantities dependent on the polarization will be denoted in the same way), $\epsilon$ is the photon energy, $c$ is the speed of light, and $h$ is the Planck constant.
 Since  the assumption on the frozen plasma is used, the bulk velocity has the only non-zero component \mbox{$v_r=-v$}, where $v$ is the absolute velocity value.

The continuity of the mass flow implies that \citep{1981A&A....93..255W}
 \beq{e:cont}
  n_{\rm e} m_{\rm p} vr_*^l=\frac{\dot M}{A}, 
 \eeq
 where $A$ is the area cut out by the accretion channel at the neutron star surface. 
 In the small-angle approximation, \mbox{$A=\pi (R \theta_0)^2$}  for a filled funnel and  \mbox{$A=\pi R^2 (\theta_2^2-\theta_1^2)$} for a hollow column.

 The  radiative transfer equations are  \citep{1981MNRAS.194.1033B, 1956Kompaneets, 1992hrfm.book.....M,Gornostaev2019}
\beqa{e:transf}
\nabla\cdot(\hat D_i\nabla n_i)-{\bm v}\cdot\nabla n_i
+\nabla\cdot{\bm v}\frac{\epsilon}{3}\frac{\partial n_i}{\partial\epsilon}\\\nonumber
+\frac{n_{\rm e}}{m_{\rm e}c\epsilon^2}\frac{\partial}{\partial\epsilon}\left(\bar\sigma_i \epsilon^4 \left(\left(kT+ \frac{m_{\rm e}v^2}{3}\right)\frac{\partial n_i}{\partial\epsilon}+n_i(1+n_i)\right) \right)\\\nonumber+n_{\rm e}c(\sigma_{i\leftarrow 3-i} n_{3-i}-\sigma_{3-i\leftarrow i}n_i)\\\nonumber
 +  \left(\frac{kT}{\epsilon}\right)^3K_i{\rm e}^{-\frac{\epsilon}{kT}}\left(1-n_i\left({\rm e}^{\frac{\epsilon}{kT}}-1\right)\right)\\\nonumber +j_{{\rm cyc,}\,i}=0, ~i=1, 2,
\eeqa 
where   $T$ is the electron temperature,  $\hat D_i$ is the diffusion tensor, $\bar\sigma_i$ is the angle-averaged scattering cross section, $\sigma_{3-i\leftarrow i}$ is the cross section of the mode switching from the mode denoted by the right-hand subscript, $m_{\rm e}$ is the electron mass,    $k$ is the Boltzmann constant,  $K_i$ is the rate of  free-free processes, and $j_{{\rm cyc},\,i}$ is the rate of production of cyclotron photons by means of de-excitations of the first Landau level because of electron-proton collisions.  (The $i$ subscript takes everywhere below the values 1 and~2.)   Apart from the mentioned processes of three latter lines of~\eqn{e:transf}, these equations thus describe the spatial diffusion, advection, first-order bulk Comptonization (which attended by the compression of the plasma flow), thermal Comptonization, and second-order bulk Comptonization. The latter is described by adding the \mbox{${m_{\rm e}v^2}/{3}$} quantity  to the electron temperature within the Doppler term in the Kompaneets operator (\citealt*{1997ApJ...488..881P,1997ApJ...487..834T}; \citealt{2012A&A...538A..67F,2016A&A...591A..29F}).  (As is known, bulk Comptonization of both kinds is associated with the Fermi processes of the non-thermal nature, `accelerating' photons.)

 In the  used coordinates, one has \mbox{$\nabla\cdot\bm v=\frac{1}{r^l}\frac{\partial}{\partial r}\left(r^l v_r\right)$} and 
\beqa{e:diffterms}
\nabla\cdot(\hat D_i\nabla n_i)= \frac{1}{r^l}\frac{\partial}{\partial r}\left( r^l D_{\|,\,i} \frac{\partial n_i}{\partial r} \right)\\\nonumber +\frac{1}{r^2 \sin\theta r_*^{l/2-1}}\frac{\partial}{\partial\theta}\left(\sin\theta D_{\perp,\,i} \frac{\partial n_i}{\partial \theta} \right),
\eeqa
where the non-zero (diagonal) components $D_{\|,\,i}$ and $D_{\perp,\,i}$ of the $\hat D_i$ tensor  describe the diffusion along and  across the magnetic field, correspondingly.
 These are represented as \mbox{$D_{\|,\,i}=c/(3n_{\rm e}\sigma_{\|,\,i})$} and \mbox{$D_{\perp,\,i}=c/(3n_{\rm e}\sigma_{\perp,\,i})$},  where $\sigma_{\|,\,i}$ and $\sigma_{\perp,\,i}$ are the corresponding transport cross sections.

 Ignoring the specific behaviour of the electron scattering cross sections near the cyclotron resonance, let us write them for the continuum below the cyclotron energy as (e.g.~\citealt{1971PhRvD...3.2303C,1974ApJ...190..141L,1987ApJ...312..666A})
  \beq{e:cs1}
  \sigma_1=\left(\frac{\epsilon}{\epsilon_{\rm cyc}}\right)^2\sigma_{\rm T}
  \eeq
   and
  \beq{e:cs2}
  \sigma_2=\left(\sin^2\alpha+\left(\frac{\epsilon}{\epsilon_{\rm cyc}}\right)^2\cos^2\alpha \right)\sigma_{\rm T},
  \eeq
   where $\epsilon_{\rm cyc}$ is the cyclotron energy at a given $r$,  $\alpha$ is the angle between the wavevector of the incoming photon and the direction of the magnetic field, and $\sigma_{\rm T}$ is the Thomson cross section.  For a medium dominated by the electron scattering (this is the case under consideration, as will be estimated and numerically verified), it is set
  \beqa{}
  \sigma_{\|,\,1}=\sigma_{\perp,\,1}=\bar\sigma_1=\sigma_{\|,\,2}=\left(\frac{\epsilon}{\epsilon_{\rm cyc}}\right)^2\sigma_{\rm T},~ \sigma_{\perp,\,2}=\sigma_{\rm T},\\\nonumber
   \bar\sigma_2=\int\limits_0^1\sigma_2 {\rm d}\cos\alpha=\frac{1}{3}\left(2+\left(\frac{\epsilon}{\epsilon_{\rm cyc}}\right)^2\right)\sigma_{\rm T}.     
\eeqa
 The mode switching cross sections are written, taking the angle-averaged value for the ordinary mode, as 
\beq{}
\sigma_{3-i\leftarrow i}= \frac{1}{4}\left(\frac{\epsilon}{\epsilon_{\rm cyc}}\right)^2\sigma_{\rm T}. 
\eeq 
 For the photon energies above the cyclotron energy, it is set that \mbox{$\sigma_{\|,\,i}=\sigma_{\perp,\,i}=\bar\sigma_i=\sigma_{\rm T}$}, and \mbox{$\sigma_{3-i\leftarrow i}=\sigma_{\rm T}/4$}. The use of a factor of one third in the spatial diffusion coefficients is a certain arbitrariness, which cannot lead to any errors in the solutions within the problem under consideration, except for an insignificant  variation in the width of the shock wave not affecting the spectrum of the funnel.

The rate of free-free interactions for the ordinary mode is well approximated  in the continuum by the non-magnetic formula \citep{1961ApJS....6..167K, 1975SvA....18..413I}
\beq{e:Kff}
K_2= \frac{8\pi}{3} \frac{e^6h^2}{\sqrt{6\pi m_{\rm e}^3}}n_{\rm e}^2(kT)^{-\frac{7}{2}}g\left(\frac{\epsilon}{kT}\right),
\eeq
where $e$ is the electron charge,  and 
\beq{}
g(x)=
\begin{cases}
1, & x\geq 1,\\
\frac{\sqrt{3}}{\pi}\ln\frac{2.35}{x}, & x<1
\end{cases}
\eeq
 is the Gaunt factor  written with the use of the Born approximation for an arbitrary $x$. 
 It is set that \mbox{$K_1=0$}, since the coefficients of the free-free absorption and emission for the extraordinary mode in the continuum are several orders of magnitude smaller compared to those for the ordinary mode at the energies of interest (e.g.~\citealt{1981ApJ...251..278N, 2022MNRAS.517.4022S}).

 Simple estimates for a non-magnetized  static medium show that the Compton scattering is a much more efficient mechanism of the energy exchange than  free-free processes in the problem under consideration.
Indeed, the corresponding rate of the Compton interactions is \citep{1975SvA....18..413I}
\beq{}
a\sim \frac{kT}{m_{\rm e}c} \sigma_{\rm T} n_{\rm e},  
\eeq
and the rate of free-free processes is (cgs) 
\beq{}
K\sim 1.2 \times 10^{-12}n_{\rm e}^2T^{-\frac{7}{2}}  
\eeq
for $\epsilon>kT$. 
In the considered parameter ranges, for both geometries $a$ significantly exceeds  $K$ throughout the column. 
 Consequently, the heating and cooling due to Compton scatterings are the main processes governing the electron temperature  within  the settling mound (post-shock zone in the velocity distribution) and the dense region of the shock wave.

The  source of cyclotron photons is given (cgs)  by the expression \citep{1987ApJ...312..666A,2007ApJ...654..435B}
\beq{e:jcyc}
\sum_{i} j_{{\rm cyc},\,i}= 1.8\times 10^{-60} \epsilon^{-2} n_{\rm e}^2 B_{12}^{-\frac{3}{2}} H\left(\frac{\epsilon_{\rm cyc}}{kT}\right) {\rm e}^{-\frac{\epsilon_{\rm cyc}}{kT}} f,
\eeq
where  \mbox{$B_{12}=B/10^{12}~{\rm G}$}, 
\beq{}
H(x)=
\begin{cases}
0.41, & x\geq 7.5,\\
0.15\sqrt{x}, & x<7.5,
\end{cases}
\eeq
and 
\beq{}
f=\delta(\epsilon-\epsilon_{\rm cyc}).
\eeq
 In the computations, the delta function is approximated by the normalized Gaussian function
\beq{}
f=\frac{1}{\sqrt{2\pi}\tilde\varsigma}\rm e^{-\frac{1}{2}\left(\frac{\epsilon-\epsilon_{\rm cyc}}{\varsigma}\right)^2},
\eeq
where $\tilde\varsigma=\varsigma/1~{\rm keV}=1$ (e.g.~\citealt{2016A&A...591A..29F,2017ApJ...835..130W}).  Normalization is formally used for the infinite range of the variable, since the distribution is centred far from zero.
 Using normalization for the semi-infinite range does not affect the results remarkably. It is assumed that \mbox{$j_{\rm cyc,\,1}=j_{\rm cyc,\,2}$}.

The local Compton equilibrium temperature (the Compton temperature) gives the value of the electron temperature in a medium when this value is determined by the spectrum of the radiation. For the case of a non-polarized radiation  in a static medium, the Compton temperature reads \citep{1970JETPL..11...35Z} 
\beq{e:temperZL}
T_{\rm C,\,np}=\frac{\int\limits_0^\infty \epsilon^4 n_{\rm np}(1+n_{\rm np}) {\rm d} \epsilon}{4k\int\limits_0^\infty \epsilon^3 n_{\rm np} {\rm d} \epsilon},
\eeq
 where $n_{\rm np}$ denotes the corresponding photon occupation number.
Multiplying this expression by the denominator, writing the resulting expression for each mode and summing equations, one can obtain
\beq{e:temper}
T_{\rm C}=\frac{\int\limits_0^\infty \epsilon^4 \sum_{i}  n_i(1+n_i) {\rm d} \epsilon}{4k\int\limits_0^\infty \epsilon^3 n {\rm d} \epsilon}.
\eeq
To describe the thermal structure of the  funnel, it can be then supposed 
 \beq{e:temper_e}
 T=\begin{cases}
 T_{\rm C}-\frac{m_{\rm e}v^2}{3k}, & T_{\rm C}-\frac{m_{\rm e}v^2}{3k}\geq T_*,\\
T_*, & T_{\rm C}-\frac{m_{\rm e}v^2}{3k}<T_*,
\end{cases}
\eeq
 where the constant $T_*$ is the electron temperature set in the region in which the temperature value cannot be fully governed by the radiation. This is the case of the pre-shock  region and the upper low-density region of the shock wave, where  too low values of the photon number density and too high values of the bulk velocity make the local Compton equilibrium approximation inapplicable. Since an exact calculation of the electron temperature in this region is not the goal of this work  and cannot change the main results, \eqn{e:temper_e} is used in the simulations. The influence of plasma compression on the temperature manifests mostly through the values of the photon occupation numbers, and the effects, which would come through the modification of the expression for the temperature, are negligible. Expressions \eqn{e:temperZL} and \eqn{e:temper} also do not include the terms responding to free-free processes.
After equating  the heating and cooling rates, it can be shown that the inclusion of these terms in \eqn{e:temperZL} and \eqn{e:temper} could affect the values of $T_{\rm C,\,np}$ and $T_{\rm C}$ at plasma densities considerably exceeding those at which the gas pressure becomes  important.

The momentum equation and the radiative transfer equations require boundary conditions. As a boundary condition for  equation~\eqn{e:mom}, it is set that
 \beq{e:vup}
v(r_{\rm up})=v_{\rm ff}(r_{\rm up})=\sqrt{\frac{2GM}{r_{\rm up}}},
\eeq
where $v_{\rm ff}$ denotes the free-fall velocity,  or $v(r_{\rm up})$ is set to be equal to a certain value. For the transfer equations, at the upper boundary it is set
\beq{e:upbound}
 n_i(r_{\rm up})=0. 
\eeq
The free escape of the photons is supposed at the sidewall (sidewalls), so that 
 \beq{e:sidebound}
 -\frac{D_{\perp,\,i}}{r}\left.\frac{\partial n_i}{\partial \theta}\right|_{\theta=\theta_0}=\frac{2}{3}cn_i(\theta_0)
 \eeq 
 for the filled funnel and
  \beq{e:sideboundh}
 (-1)^d\frac{D_{\perp,\,i}}{r}\left.\frac{\partial n_i}{\partial \theta}\right|_{\theta=\theta_{d+1}}=\frac{2}{3}cn_i(\theta_{d+1}),~~d=0,~1
 \eeq 
 for the hollow funnel.
 It is supposed that the blackbody spectrum is established at the bottom, so that 
 \beq{e:botbound}
 n_i(R)=\frac{1}{2}\left({\rm e}^{\frac{\epsilon}{kT_{\rm 0}}}-1\right)^{-1},
 \eeq
  where 
  \beq{e:bottemp}
  T_{\rm 0}= \left( \frac{3 \gamma \dot M v(r_{\rm up})}{A a_{\rm r}}\right)^\frac{1}{4}, 
  \eeq
  with $a_{\rm r}$ is the radiation energy density constant and  
\mbox{$\gamma\approx 1-v(R)/v(r_{\rm up})\sim 1$} is the parameter (in the uniform magnetic field \mbox{$\gamma = 1-v(R)/v(r_{\rm up})$}).  For the filled column 
\beq{e:axis}
\left.\frac{\partial n_i}{\partial \theta}\right|_{\theta=0}=0.
\eeq 
 At the boundaries  of the photon energy domain corresponding to \mbox{$\epsilon_1= 0.07$~keV} and \mbox{$\epsilon_2= 500$~keV} it is set that 
 \beq{e:epsbound}
 n_i(\epsilon_1)=n_i(\epsilon_2)=0.
 \eeq

Regarding the boundary conditions for the transfer equations, it should be noted that the top boundary is different from the side ones, since there is no any real physical boundary of the medium at the top (the structure of the flow up to  the magnetospheric boundary is not discussed here).      
Relatively low-intensity spectrum taking place not so far from that boundary cannot be simply an additive contribution to the outgoing radiation.  A lot of the photons  forming that spectrum -- sooner or later -- get knocked down and thus became a contribution to the sidewall emission, which is determined by solving the spatially 2D radiative transfer under the conditions of the funnel. The radiation flux along the magnetic field vanishes at large $r$. The spectrum of radiation emerging from the channel  above the set upper boundary can be found by modelling the radiative transfer in the corresponding spatial regions.

The substantially non-stationary situation near the column base cannot be simulated in detail while one is constrained by the considered system of equations, therefore the values of the electron number density obtained for the corresponding region may differ from real those.

\subsection{Problem parameters}
\label{ssec:param}

 The value of \mbox{$\dot M_{17}=\dot M/10^{17}~{\rm g~s^{-1}}$} is varied in the range~\mbox{$0.2$--$5$}.  
It is always set that \mbox{$\epsilon_{\rm cyc}(R)=50$~keV}.

The radius of the funnel base depends on the accretion rate. The angular half-width of the funnel base \mbox{$\theta_0(\dot M_{17}=1)$} is taken as a parameter of the problem for a filled funnel, while \mbox{$\theta_1(\dot M_{17}=1)$} and \mbox{$\theta_2(\dot M_{17}=1)$} are parametrized for a hollow funnel. It is supposed that \mbox{$\theta_2=1.1\theta_1$} and \mbox{$\theta_1=\theta_0$}, with \mbox{$\theta_0=\theta_0(\dot M_{17}=1)\dot M_{17}^{1/7}$} \citep*{1973ApJ...184..271L, 1983bhwd.book.....S}. The angular width of the funnel base is not strictly related to the value of the surface magnetic field.  It is set that \mbox{$\theta_0(\dot M_{17}=1)= 0.08$}  (one would exclude the parameters $\theta_0$, $\theta_1$ and $\theta_2$,  and set them as some functions of $B_{\rm ns}$ and $\dot M$ for a given neutron star).

 It is set that \mbox{$M=1.4M_\odot$} and  \mbox{$R=10^6$~cm}. In what follows, the case of \mbox{$l=3$} corresponding to the dipole magnetic field is considered.  It is supposed that \mbox{$T_* = 1~{\rm keV}$}, which is of the order of magnitude equal to the real electron temperature in the flow  above the emitting region. It is set that \mbox{$\gamma=0.98$}. The  $v(r_{\rm up})$ quantity is considered as a parameter in the particular case, where \mbox{$v(r_{\rm up})=1.3\times 10^{10}$~cm~s$^{-1}$}.

 Higher values of the mass accretion rate as well as distinct values of the surface magnetic field strength, the radius of the funnel base and the thickness of the filled part of the hollow funnel   will be considered in a separate paper.

\subsection{Numerical approach}

Computations are performed with the use of the time-relaxation method. The initial spectrum in the medium is the Wien distribution for the temperature of \mbox{$1$~keV}. At each time step, to find the velocity with the use of \eqn{e:mom} at a given point in space for some spectral distribution, the total radiation energy density is determined  by 
 expression~\eqn{e:dens}. The electron number density is eliminated in \eqn{e:mom} with the use of~\eqn{e:cont}. After integrating the momentum equation, the electron temperature is found with the use of  \eqn{e:temper_e}  over the entire computational domain (the Stefan-Boltzmann law is not used anywhere except for calculating the value of~$T_0$).

To find the photon occupation number at each time step, an explicit  finite-difference scheme developed by G2021 is  used, with additions and modifications related to the physical picture of the problem and  aimed at the approximation of the radiative transfer equations \eqn{e:transf} in their specific form considered here.  There is no the sense to write out the dimensionless form of the equations,  since the example of such a representation can be found by the reader in the mentioned paper  (the grids are similar with accuracy to the steps used). For a given $i$, the term  describing thermal and second-order bulk Comptonization is  rewritten after performing differentiation as a sum of terms, each of which contains a derivative of the dimensionless spectral radiation energy density of only one order or does not contain its derivatives of an order higher zeroth. These terms are approximated, as well as the advection and first-order bulk Comptonization terms, according to the typical rules \citep{1986nras.book.....P, 2001samarskii_eng, Kalitkin}.    The terms containing the derivatives of the cross sections $\bar\sigma_i$ with respect to $\epsilon$ are taken into account.  In each transfer equation, the $n_i^2$ nonlinearity originates three monomials, one of which contains the product of the dimensionless spectral radiation energy density and its first derivative approximated by the central difference, and the others contain the specified function squared.  To approximate the spatial diffusion terms (see~\eqn{e:diffterms}), the pattern functionals containing the arithmetic means \citep{1962samarskii} are basically used  -- the results are nearly identical to those obtained by means of the approximation after differentiation of the brackets.

The codes for simulating the accretion columns were written in C and executed in parallel with the use of OpenMP. The calculations were performed using total capabilities of two \mbox{56-core}  (\mbox{AMD EPYC 7663}) CPUs. The overall compute time of the  presented simulations is about three months.

 \section{Results } \label{sec:res} 

\subsection{Accretion column in the radiation-dominated regime}
\label{sec:rcol}

The dynamical picture of the problem is known. Consider a freely falling flow. Let \mbox{X-ray} photons be generated at the base of the funnel, and let their total luminosity be determined by the expression \eqn{e:Laccr} (a separate consideration of the case when  photons are produced in the region of a certain height above the atmosphere is omitted). An increase in the accretion rate and the \mbox{X-ray} luminosity causes an increase of the role of the pressure of the radiation propagating upwards. The flow begins to slow higher up, and the \mbox{X-ray} photons escaping the funnel begin to partially gain the energy in the appearing region of small values of \mbox{$|\nabla\cdot\bm v|$}, and even somewhat above it, in the near-divergence-free flow. A further increase in the accretion rate provides an increase in the fraction of photons gaining considerable energy in the shock wave and thus an increasingly sharp convergence of the flow. Once a sufficiently strong shock wave is formed,  the settling mound begins to arise, as a significant portion of the photons penetrate  beneath the shock. The seed free-free radiation is always originated mostly inside the underlying atmosphere.  Photons appeared inside the settling mound accomplish random walks before escaping after a time equal on average to the mean diffusion time for the mound. As the accretion rate increases further,  the shock wave becomes steeper and  the peak height of the deceleration zone grows. Both effects occur (up to achieving some state), since the profile of the shock wave must provide the illumination of the radiation for a given accretion luminosity.   A relatively small part of the pre-shock kinetic energy of protons is converted into the thermal energy of electrons.
  Within the shock, there is a surface where the supersonic decelerating flow becomes  subsonic (the speed of sound in a radiation-dominated plasma is \mbox{$c_{\rm s}=2/3\sqrt{u/(m_{\rm p} n_{\rm e})}$}).

The spectrum of the radiation in the column is almost entirely determined by the Compton energy exchange between  electrons and photons.
Modelling the radiative transfer in the hot rarefied plasmas in the presence of a strong magnetic field leads to  results for the emergent spectrum distinguishing from those for a nonmagnetized medium, mainly in the consequence of the dependence of the differential scattering cross sections  on polarization, frequency, directions of propagation of incoming and outgoing photons, magnetic field strength, electron temperature and  mass density \citep{1992hrfm.book.....M}. The dependence of the cross sections on the electron temperature is of importance mostly for modelling the radiative transfer in the vicinity of cyclotron resonances, whose broadening is more remarkable, the higher the electron temperature. The Compton \mbox{$y$-parameter} characterising thermal Comptonization in the continuum 
(e.g.~\citealt{1979rpa..book.....R})  is reduced considerably compared to the nonmagnetic value for both modes  at frequencies well less than the cyclotron frequency.  Even for  values of the nonmagnetic \mbox{$y$-parameter} much exceeding unity, under the conditions of the hard thermal backlight and static isothermal medium, magnetic Comptonization can cause an emergent spectrum summed over the modes, which has a nearly power-law shape at high frequencies.   The general description of Comptonization in strongly magnetized plasmas is rather complicated and remains beyond the scope of this paper.

Geometrically 2D models of the radiation-dominated accretion column differ from 1D models by that the 2D model settling zone is surrounded by the shock wave and, thus, completely submerged in the flow, so that there is no  direct path between that zone and the outer space.  
Such models should be of interest in light of finding their relationship    to  observational data.

  The model developed here is applicable for values of the parameters which provide at least the very beginning of originating the settling zone.
For the total luminosity of two identical columns one can write that
\beq{e:Lx}
L_{\rm X}\sim\dot Mv^2(r_{\rm up}).
\eeq
For the case of a freely falling pre-shock flow one has \mbox{$L_{\rm X}\approx 2L$}.
At reasonable choice of \mbox{$r_{\rm up}$}  expression \eqn{e:Lx} characterises approximately, in principle, any emitting region regardless that whether it being in the radiation-dominated state or not.  The specific of the radiation-dominated column is that the temperature $T_0$ is relatively higher than calculated  for the same parameters the effective temperature of the hot spot
\beq{e:Tth}
T_{\rm eff}\approx\left(\frac{\dot M v^2(r_{\rm up})}{2Ab}\right)^\frac{1}{4},
\eeq
where $b$ is the Stefan-Boltzmann constant. The ratio of the temperatures is  
\beq{e:T0th}
\frac{T_0}{T_{\rm eff}}\approx\left(\frac{3}{2}\frac{c}{v(r_{\rm up})}\right)^\frac{1}{4}.
\eeq
That is, the processes characteristic for the shock wave provide  more effective heating of the medium than that would take place in a `pure thermal' atmosphere. 


\subsection{Results of the numerical calculations}
\label{sec:numres}

\begin{figure*}
 	\begin{center}
     \includegraphics[width=0.19\textwidth]{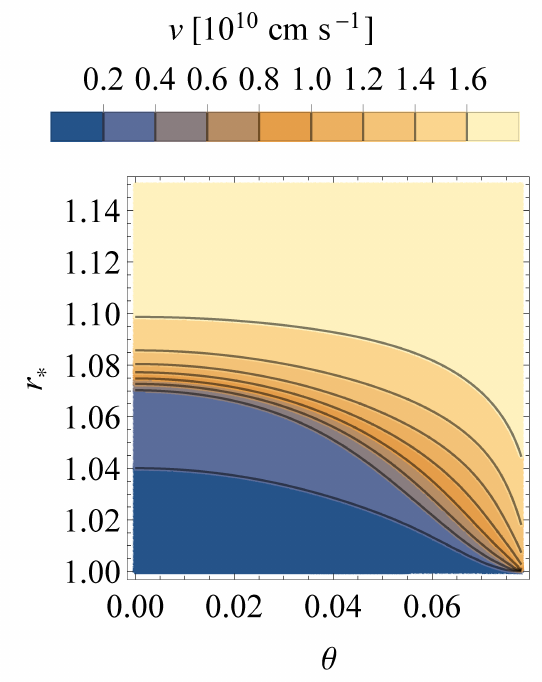}
  \hfill
  \includegraphics[width=0.19\textwidth]{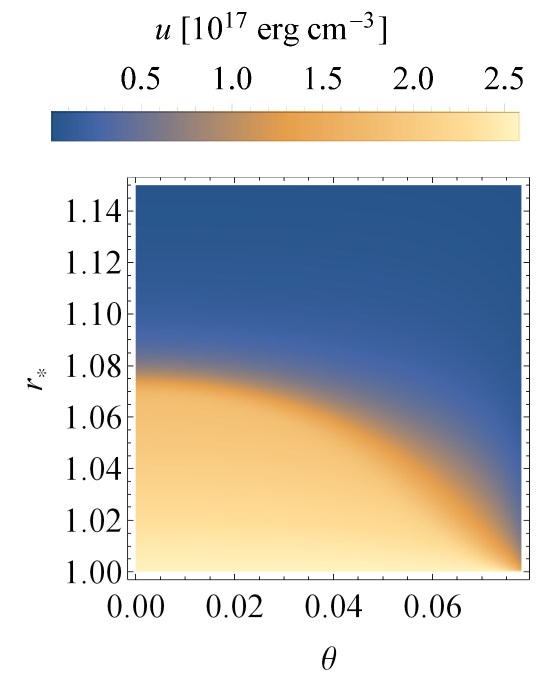}
    \hfill
    \includegraphics[width=0.19\textwidth]{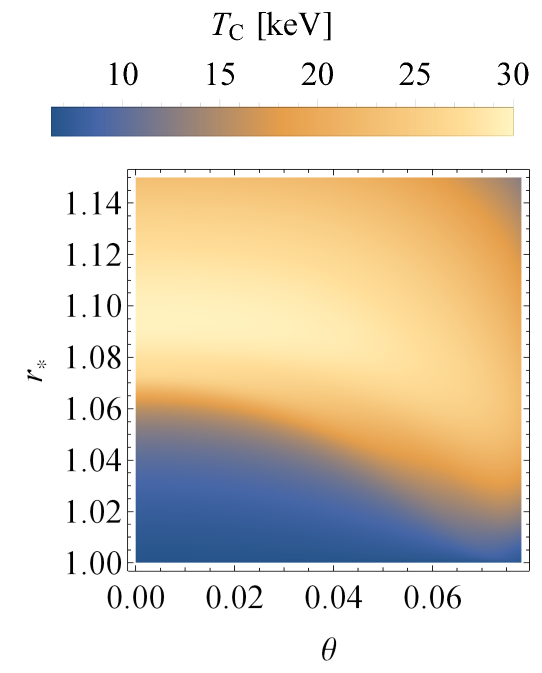}
        \hfill
   \includegraphics[width=0.19\textwidth]{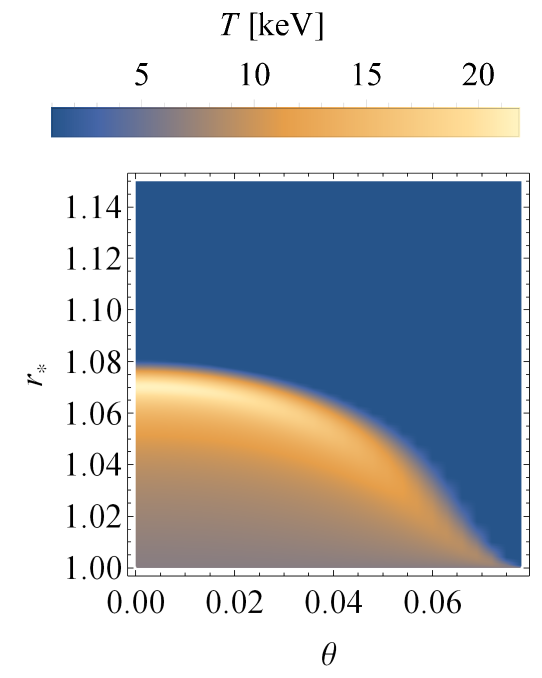}
          \hfill
   \includegraphics[width=0.19\textwidth]{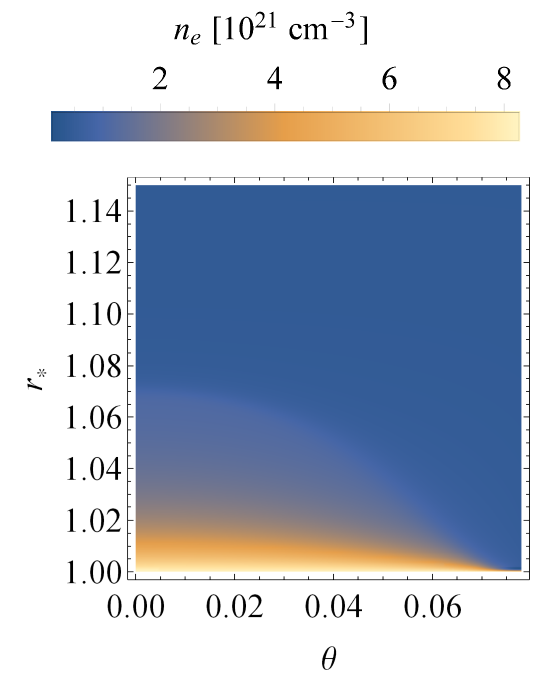}\\\vspace{2mm}
   \includegraphics[width=0.48\textwidth]{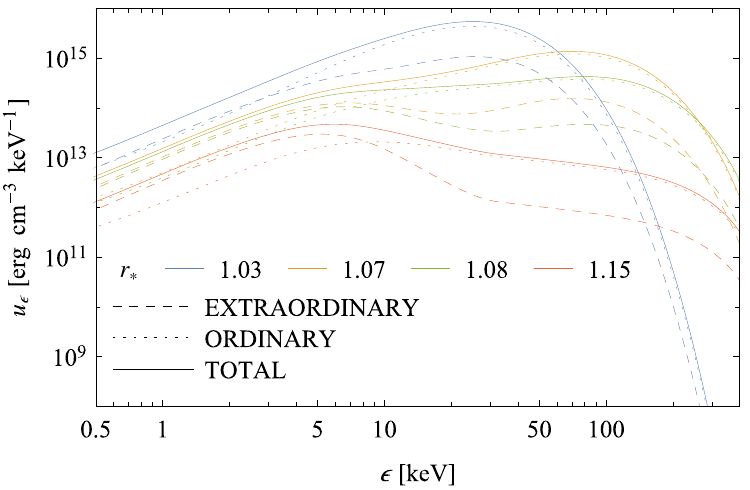}
  \hfill
  \includegraphics[width=0.48\textwidth]{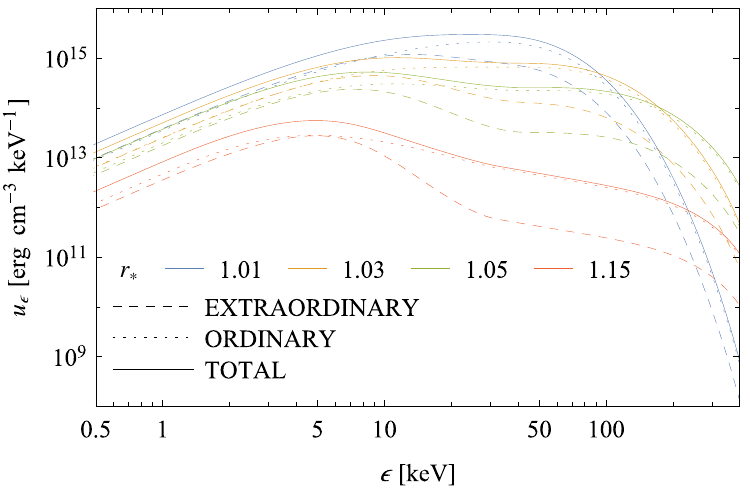}
  \\\vspace{2mm}
   \includegraphics[width=0.48\textwidth]{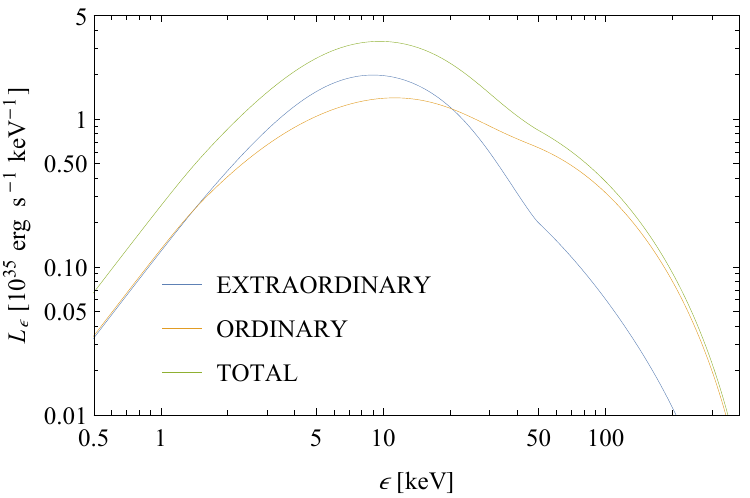}
  \hfill
  \includegraphics[width=0.48\textwidth]{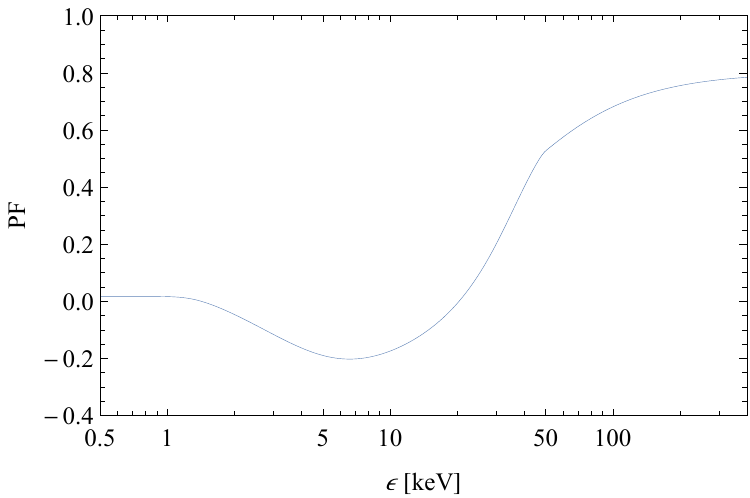}
  		\caption{Column grown in the filled accretion funnel, calculations for \mbox{$\dot M_{17}=1$}: (top panels) the structure, (middle)  the spectrum of the radiation in the number of points within the funnel, (left) \mbox{$\theta=0$}  and  (right) \mbox{$\theta=0.07$}, (bottom left) the spectral luminosity of the radiation  from the side boundary of the funnel, and (bottom right) the polarization fraction for the emergent radiation.
  		}
  \label{fig:str1}
	\end{center}
 \end{figure*}

\begin{figure*}
 	\begin{center}
     \includegraphics[width=0.19\textwidth]{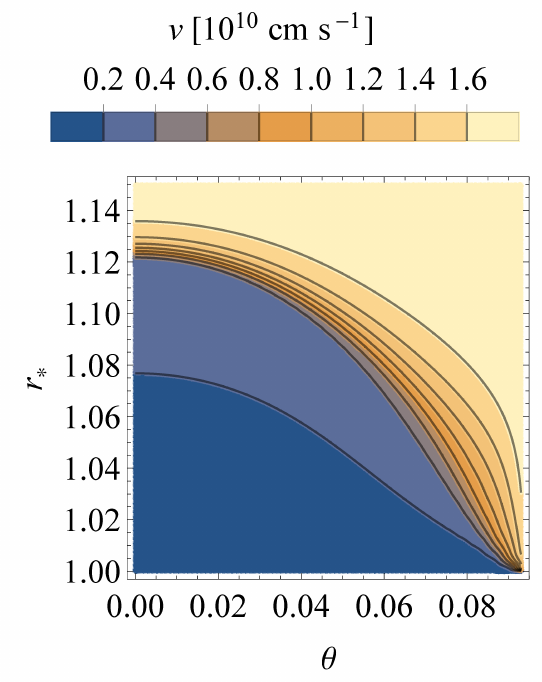}
  \hfill
  \includegraphics[width=0.19\textwidth]{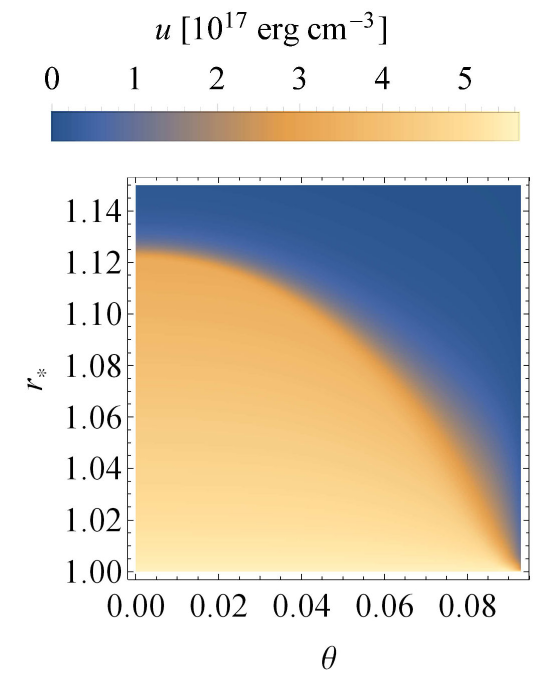}
    \hfill
    \includegraphics[width=0.19\textwidth]{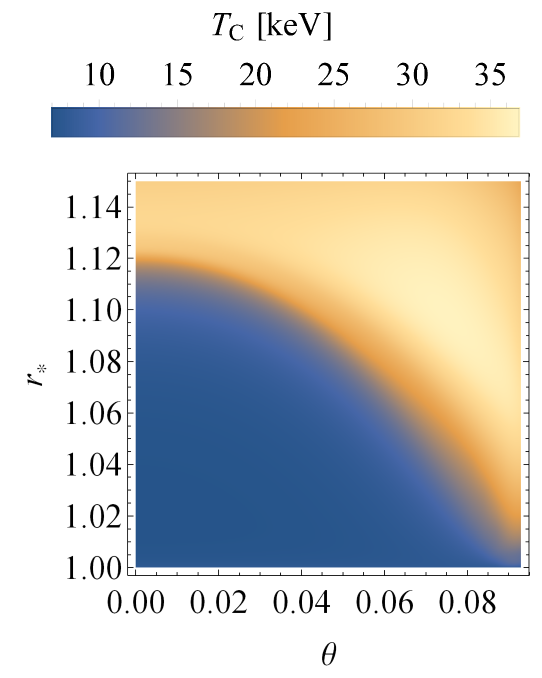}
        \hfill
   \includegraphics[width=0.19\textwidth]{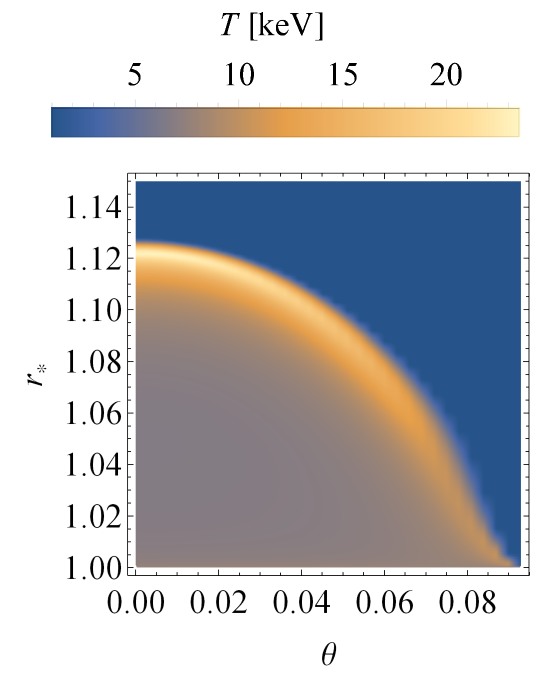}
          \hfill
   \includegraphics[width=0.19\textwidth]{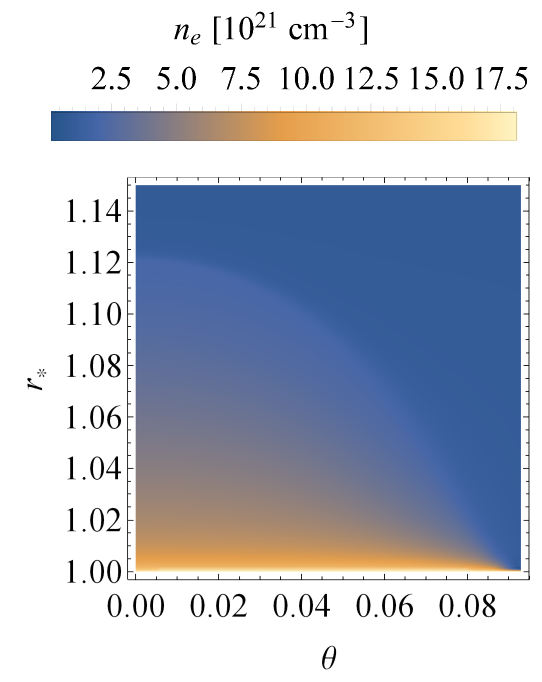}\\\vspace{2mm}
   \includegraphics[width=0.48\textwidth]{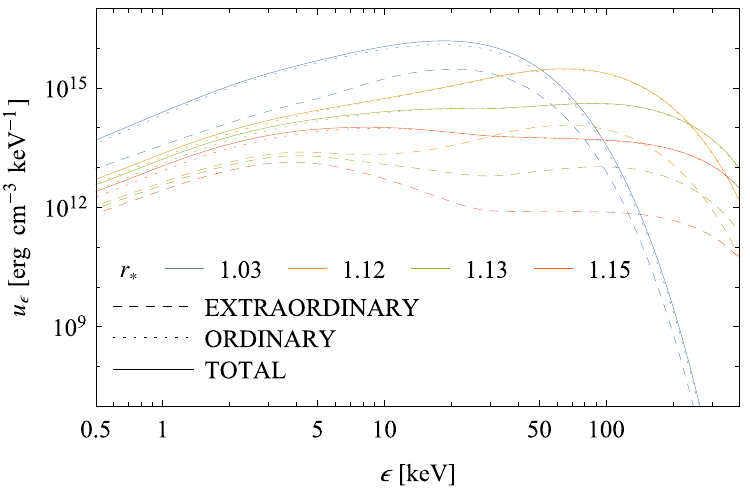}
  \hfill
  \includegraphics[width=0.48\textwidth]{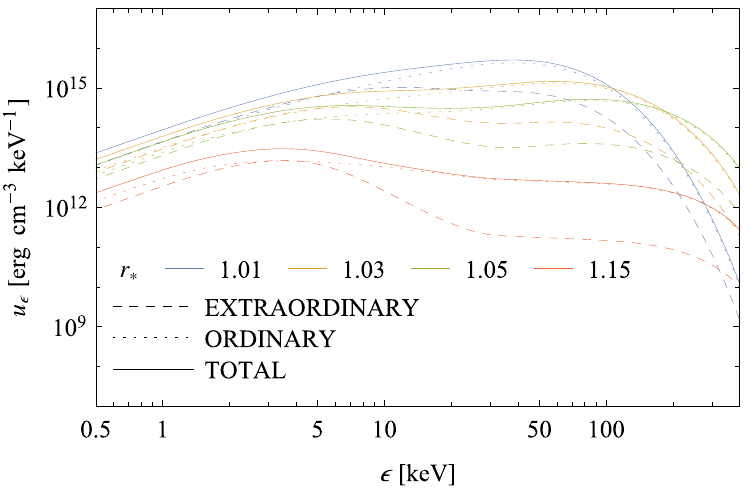}
  \\\vspace{2mm}
   \includegraphics[width=0.48\textwidth]{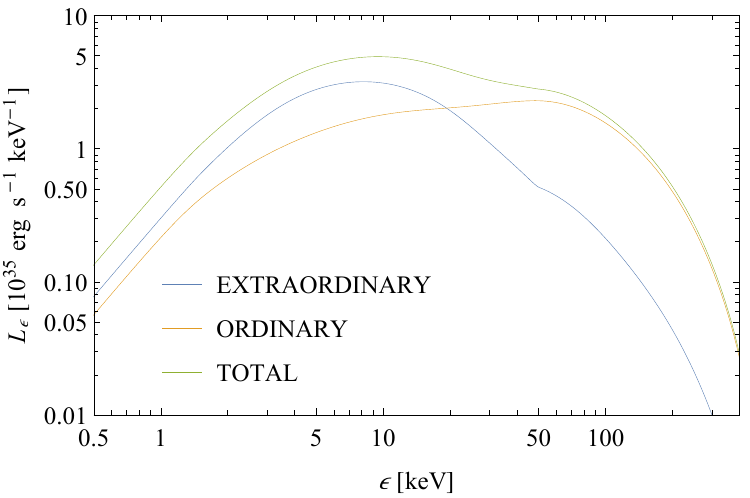}
  \hfill
  \includegraphics[width=0.48\textwidth]{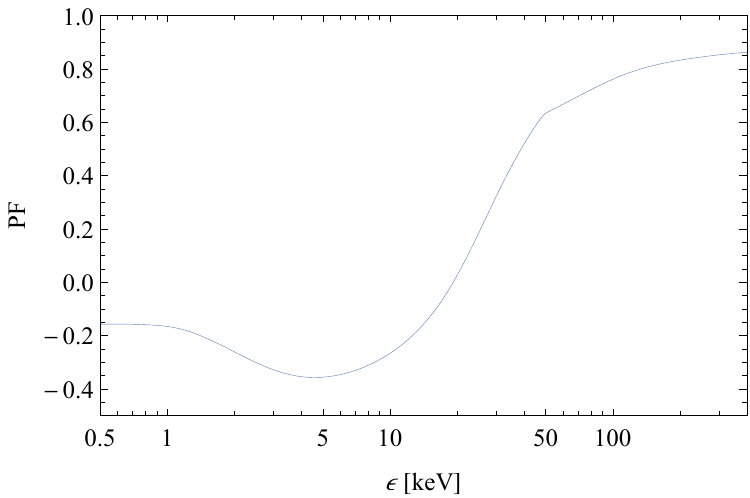}
  		\caption{Same as Fig. \ref{fig:str1}, but for $\dot M_{17}=3$. In the right panel, the spectral radiation energy density is shown for \mbox{$\theta=0.087$}.
  		}
  \label{fig:str3}
	\end{center}
 \end{figure*}

\begin{figure*}
 	\begin{center}
     \includegraphics[width=0.19\textwidth]{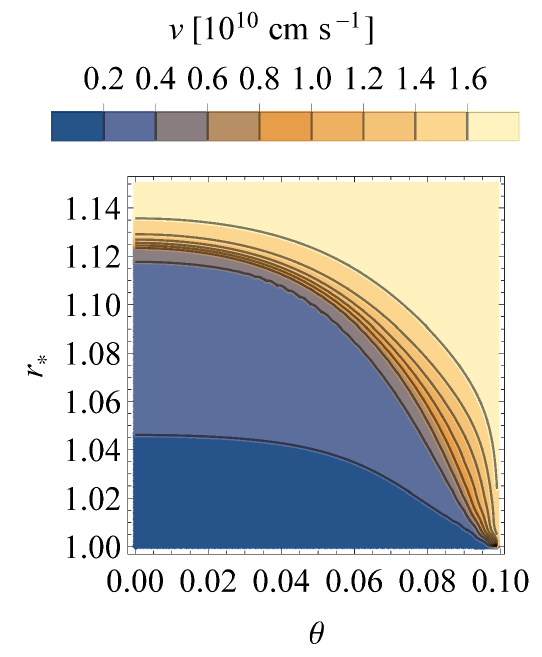}
  \hfill
  \includegraphics[width=0.19\textwidth]{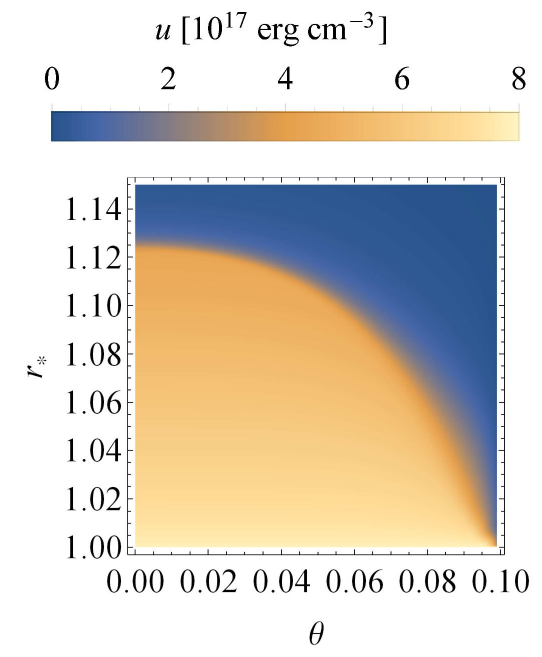}
    \hfill
    \includegraphics[width=0.19\textwidth]{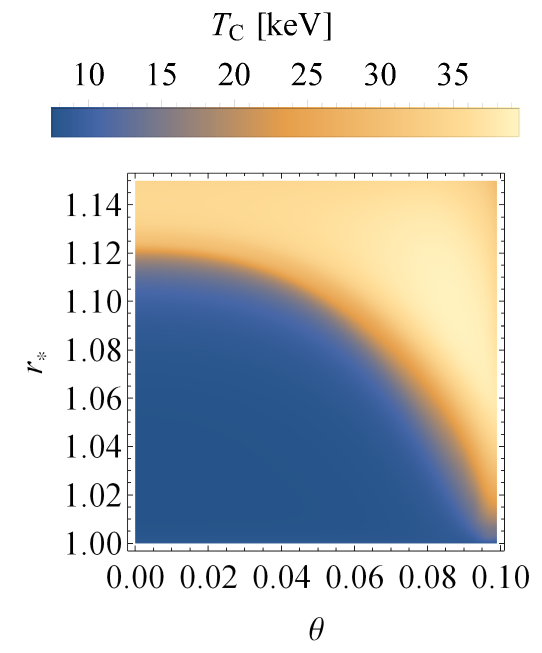}
        \hfill
   \includegraphics[width=0.19\textwidth]{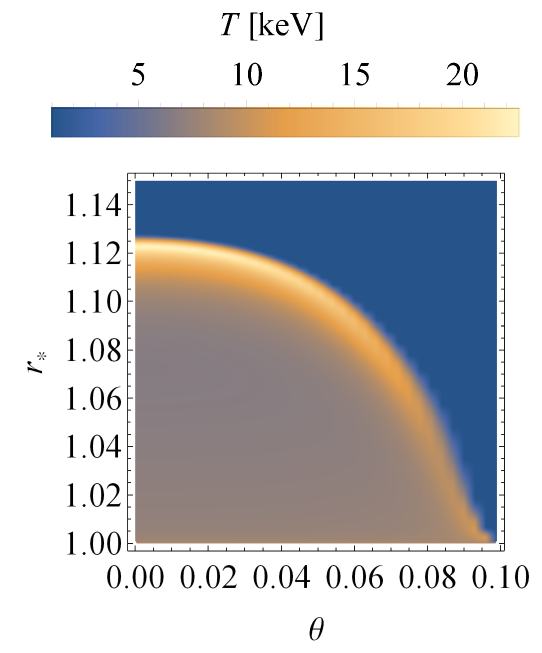}
          \hfill
   \includegraphics[width=0.19\textwidth]{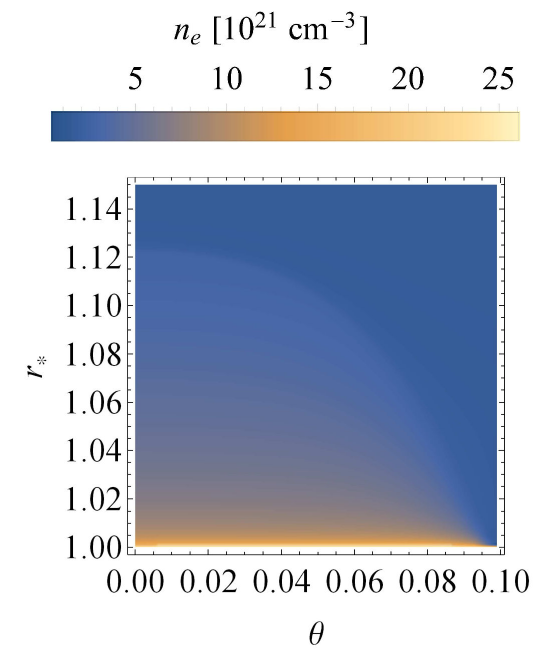}\\\vspace{2mm}
   \includegraphics[width=0.48\textwidth]{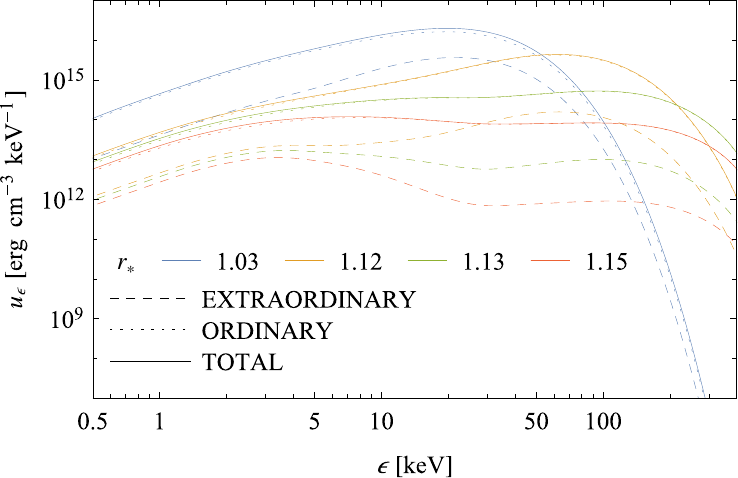}
  \hfill
  \includegraphics[width=0.48\textwidth]{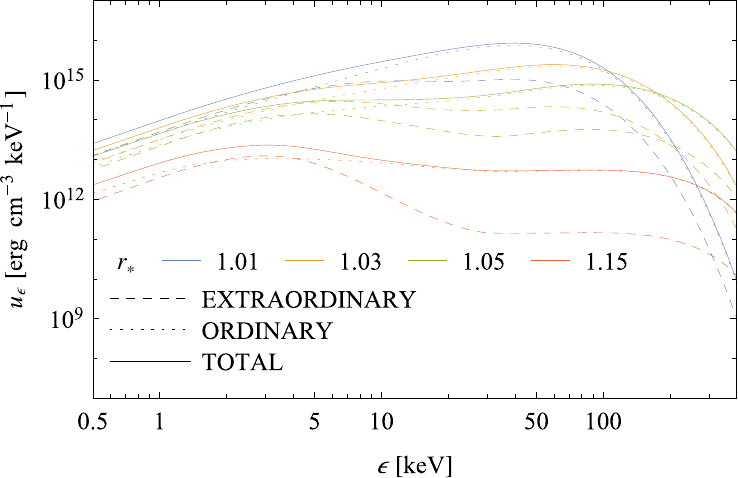}
  \\\vspace{2mm}
   \includegraphics[width=0.48\textwidth]{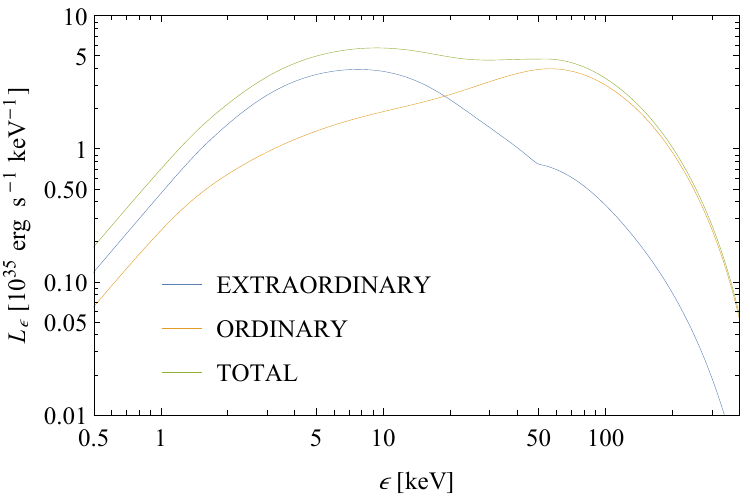}
  \hfill
  \includegraphics[width=0.48\textwidth]{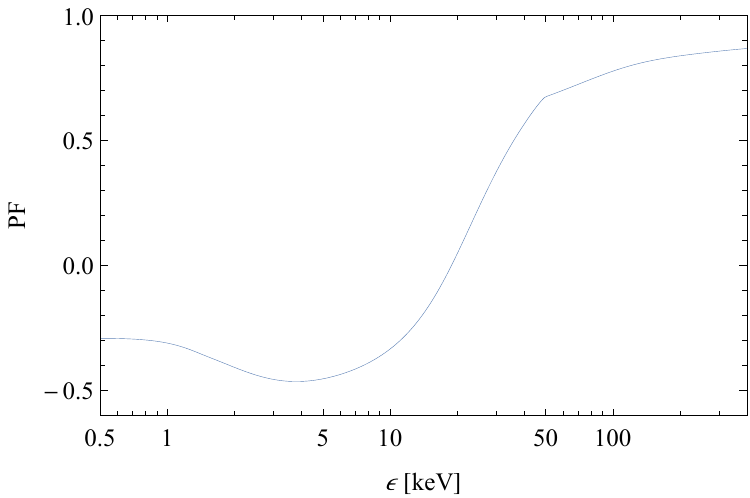}
  		\caption{Same as Fig. \ref{fig:str1}, but for $\dot M_{17}=5$. In the right panel, the spectral radiation energy density is shown for \mbox{$\theta=0.093$}.
  		}
  \label{fig:str5}
	\end{center}
 \end{figure*}

\begin{figure*}
 	\begin{center}
     \includegraphics[width=0.19\textwidth]{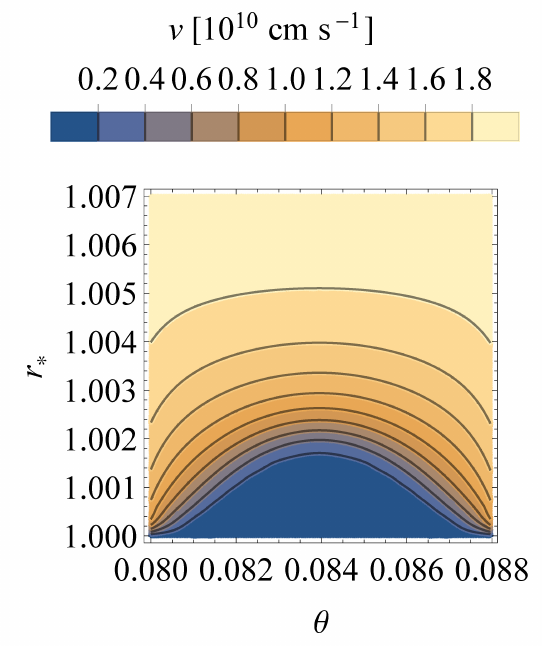}
  \hfill
  \includegraphics[width=0.19\textwidth]{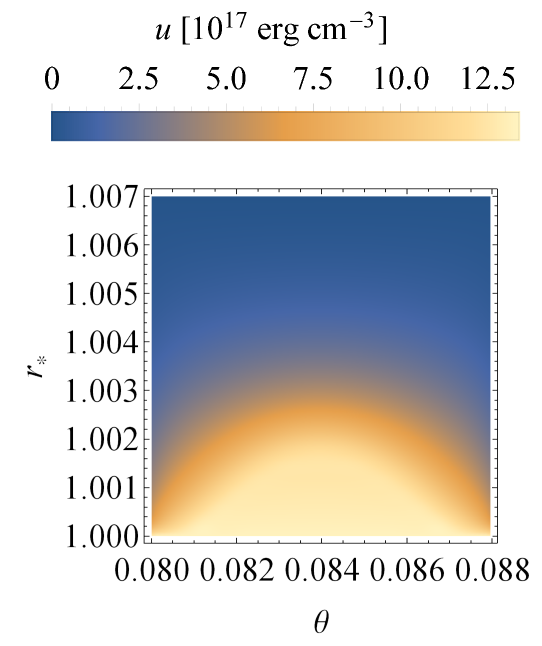}
    \hfill
    \includegraphics[width=0.19\textwidth]{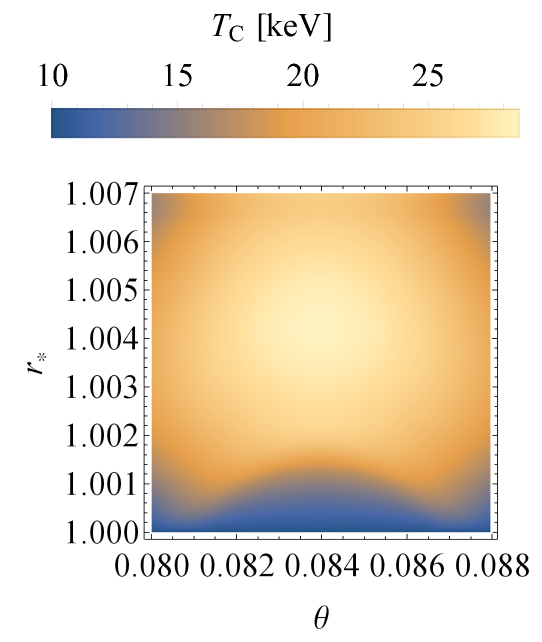}
        \hfill
   \includegraphics[width=0.19\textwidth]{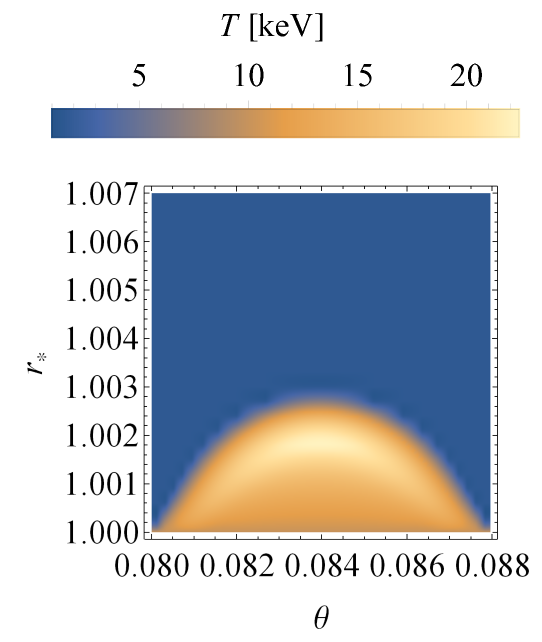}
          \hfill
   \includegraphics[width=0.19\textwidth]{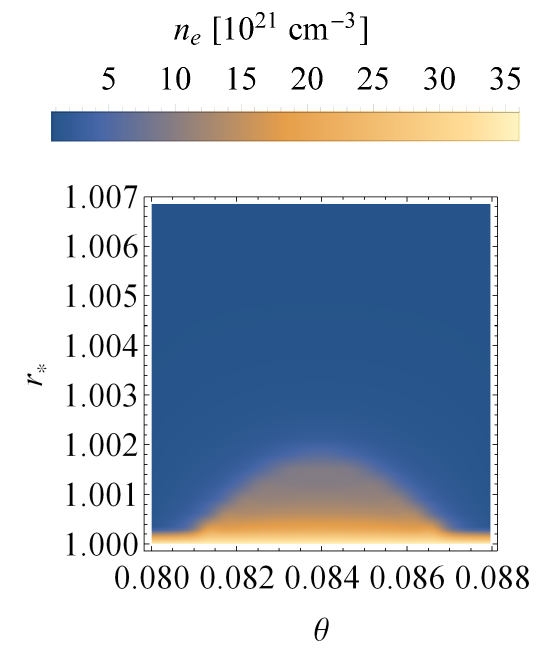}\\\vspace{2mm}
   \includegraphics[width=0.48\textwidth]{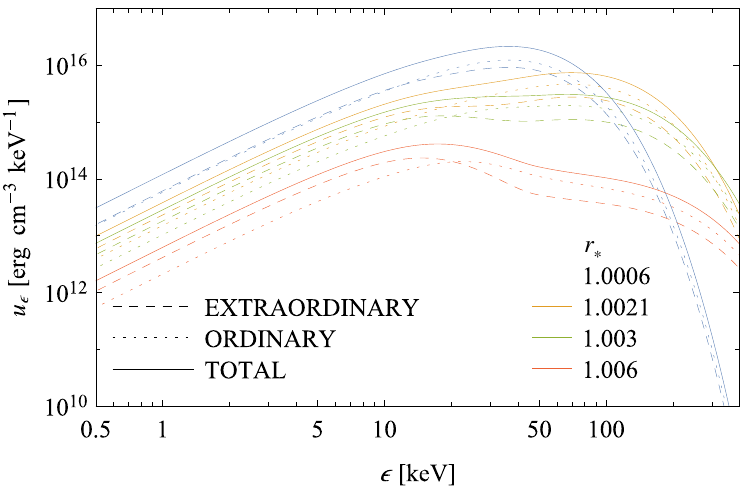}
  \hfill
  \includegraphics[width=0.48\textwidth]{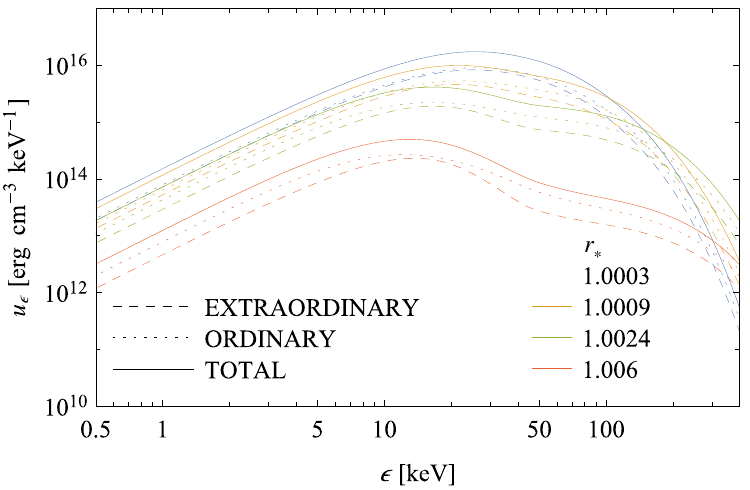}
  \\\vspace{2mm}
   \includegraphics[width=0.48\textwidth]{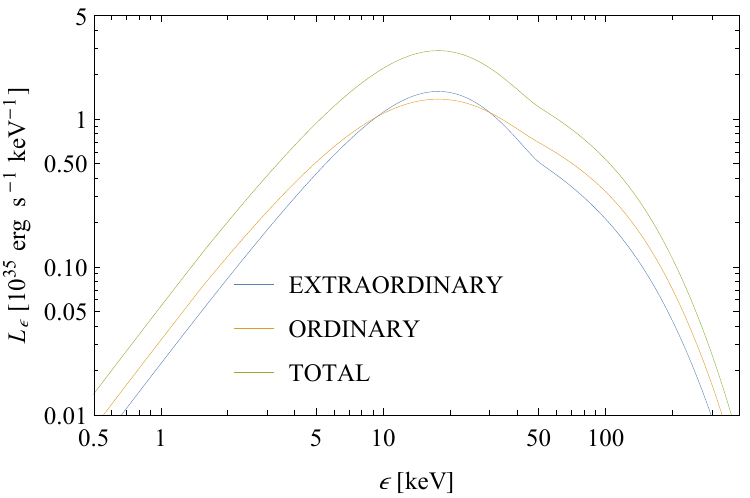}
  \hfill
  \includegraphics[width=0.48\textwidth]{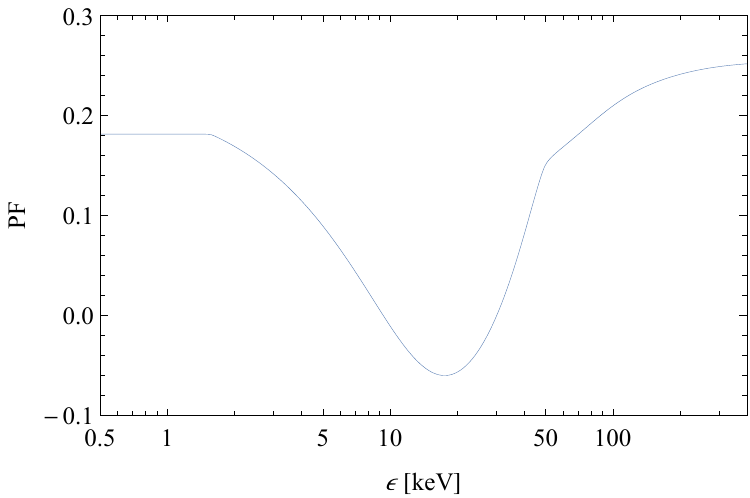}
  		\caption{Same as Fig. \ref{fig:str1}, but for the hollow funnel. The spectral  radiation energy density is shown for (left) \mbox{$\theta\approx (\theta_1+\theta_2)/2$} and (right) \mbox{$\theta\approx 0.0875$}.
  		}
  \label{fig:strh}
	\end{center}
 \end{figure*}

The  solutions to system of equations \eqn{e:mom}, \eqn{e:dens}, \eqn{e:cont}, \eqn{e:transf} and \eqn{e:temper_e} with boundary conditions \eqn{e:vup}, \eqn{e:upbound}, \eqn{e:sidebound}, \eqn{e:botbound}, \eqn{e:axis} and \eqn{e:epsbound} (or \eqn{e:sideboundh} instead of both \eqn{e:sidebound} and \eqn{e:axis}) are considered.  The main results of the current simulations are shown  in~\mbox{Figs~\ref{fig:str1}--\ref{fig:strh}}.  
 The local spectrum throughout the column is plotted in terms of the spectral radiation energy density in some points. 
The spectral luminosities $L_\epsilon$, $L_{\epsilon,\,i}$ represent the sidewalls-integrated emergent continuum, and the polarization fraction is defined as
\beq{e:pf}
{\rm PF}=\frac{L_{\epsilon,\,2}-L_{\epsilon,\,1}}{L_{\epsilon,\,1}+L_{\epsilon,\,2}}. 
\eeq

The boundaries of the colour bars related to the 2D distributions approximately indicate the minimum and maximum values approached within the  shown domains (for the velocity, these values fall within the corresponding boundary intervals). The difference between $\theta_0$, $\theta_1$ and $\theta_2$ in calculations and their values indicated above lies within the grid steps.  Seediness at the front of the shock wave in the electron temperature distributions is the trace of the colour interpolation and is neither a manifestation of the scheme instabilities nor a property of the grid functions.

 For a filled funnel \mbox{$r_*(r_{\rm up})=1.2$} (\mbox{$L_{\rm X,\,37}\approx 3.1$} for \mbox{$\dot M_{17}=1$}, \mbox{$L_{\rm X,\,37}\approx 9.3$} for  \mbox{$\dot M_{17}=3$},  and \mbox{$L_{\rm X,\,37}\approx 15.6$} for  \mbox{$\dot M_{17}=5$}), where \mbox{$L_{\rm X,\,37}=L_{\rm X}/10^{37}$~erg~s$^{-1}$}. In the case of a hollow funnel \mbox{$r_*(r_{\rm up})=1.007$} (\mbox{$L_{\rm X,\,37}\approx 3.7$} for \mbox{$\dot M_{17}=1$}). The luminosity is given here for \mbox{$v(r_{\rm up})=v_{\rm ff}(r_{\rm up})$}. The spectral luminosity shown below for a hollow funnel  is calculated as the sum of the luminosities of the side boundaries. The labels `$u_{\epsilon,\,i}$' and `$L_{\epsilon,\,i}$' are dropped in the figures.

 The self-consistent approach used allows  to simulate a specific profile of the shock wave corresponding to the frequency dependence of the cross sections for a given \mbox{$B(r)$} and to the local spectrum formed.  The  calculated characteristic values of \mbox{$\nabla \cdot \bm v$} correspond approximately to those obtained in the grey approximation for rather large cross section along the magnetic field \mbox{$\sim 0.3\sigma_{\rm T}$--$\sigma_{\rm T}$} (e.g.~\citealt{1973NPhS..246....1D}).  The growth of the column slows down as the accretion rate increases in the considered range of its values mainly due to the frequency dependence of the cross sections along the magnetic field.  The growth  virtually comes out to the saturation taking place at $\dot M_{17}$ in the range~\mbox{$3$--$5$} \mbox{(Figs~\ref{fig:str3} and \ref{fig:str5})}.  
 The vertical widening of the flow in the non-uniform magnetic field, and the effects of the frequency photon redistribution and the heating of the plasma always make the dependence of the height of the mound on the accretion rate less steep compared to the simple solutions of \cite{2015MNRAS.452.1601P},  but do not lead to its strong slowing down at \mbox{$\dot M_{17}\gtrsim 1$} being taken into account under the grey approximation.

The polarimetric spectrum demonstrates typically an excess of the radiation of the extraordinary mode at energies  near the maximum of the spectrum of that mode and prevailing the ordinary mode at high energies. The dependence of the radiation flux on the photon energy at low energies \mbox{$\propto \epsilon^2$} is always  the same for both modes.  
 At low accretion rates, the emergent spectra of both modes have the maximums at energies not exceeding strongly 
 the value~\mbox{$\sim kT_0$}. 
  With increasing accretion rate, the shape of spectrum of the extraordinary mode varies insignificantly at low energies with hardening the tail, while the spectrum of the ordinary mode gradually tends to saturation. 
 The high-energy tails are formed by the effects of bulk motion and  thermal Comptonization. 
 While the spectrum of the ordinary mode is quite far  from saturation, the frequency behaviour of the used cross sections causes that the total spectrum falls steeply below the `averaged' cyclotron line energy which thus manifests in the absence of resonance in the cross sections.  In the spectrum of the extraordinary mode that feature is held 
 for any~$\dot M$. It should be modified in exactly calculated spectra by the cyclotron line and wings of the line.

The growth of the accretion rate from \mbox{$\dot M_{17}=1$} to \mbox{$\dot M_{17}=5$} causes a decrease in the spectral hardness calculated as the ratio of the luminosity in the range \mbox{$5$--$12$~keV}  to luminosity in the range \mbox{$1$--$3$~keV}   with increasing $\dot M$ with going out to an almost constant.\footnote{The similar energy ranges were used in the analysis of \cite{2015MNRAS.452.1601P}. The minimum and maximum energies are close to the boundaries of the {\it RXTE}/ASM energy band.}    
 A softening occurs as the relative contribution of the ordinary mode to the luminosity decreases in the range \mbox{$5$--$12$~keV}  faster than in the range \mbox{$1$--$3$~keV} during a tending to the saturation of the  spectrum of the ordinary mode  (a hardening of the total spectral tail) at increasing $\dot M$ and fixed other parameters.

\begin{figure*}
 	\begin{center}
     \includegraphics[width=0.19\textwidth]{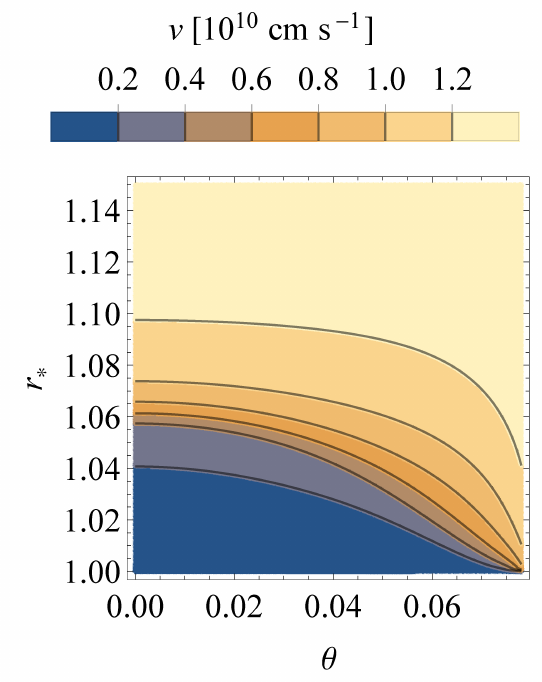}
  \hfill
  \includegraphics[width=0.19\textwidth]{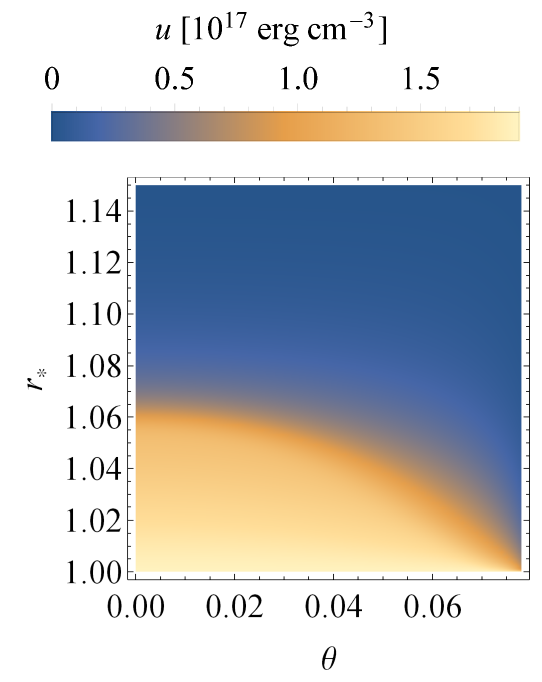}
    \hfill
    \includegraphics[width=0.19\textwidth]{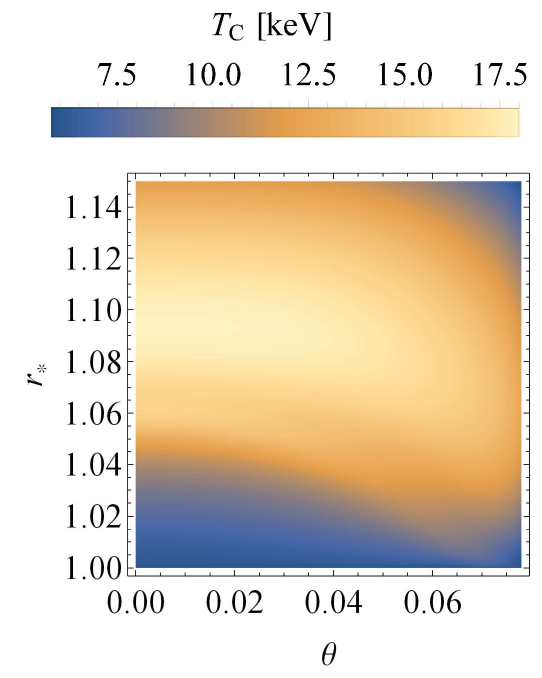}
        \hfill
   \includegraphics[width=0.19\textwidth]{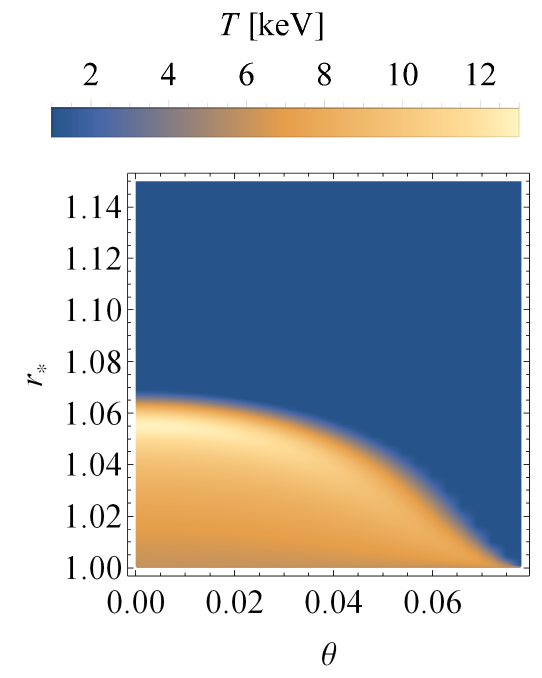}
           \hfill
   \includegraphics[width=0.19\textwidth]{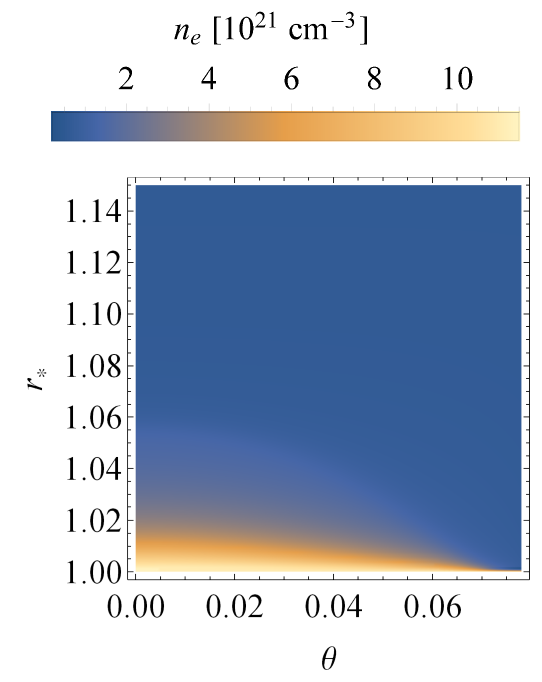}
  \\\vspace{2mm}
   \includegraphics[width=0.48\textwidth]{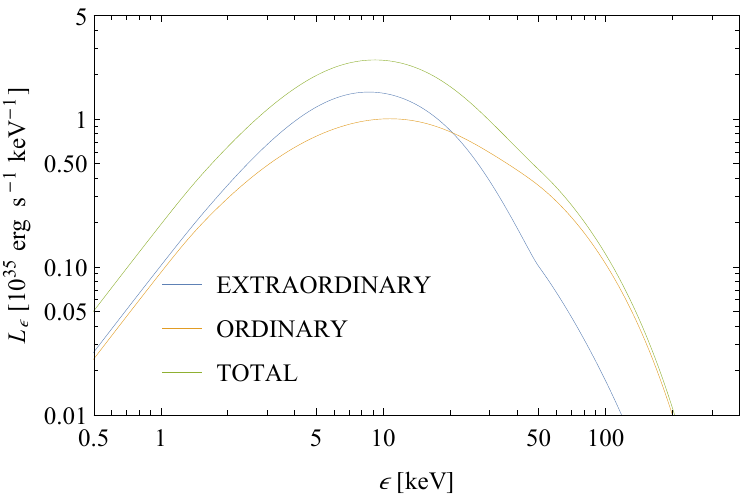}
  \hfill
  \includegraphics[width=0.48\textwidth]{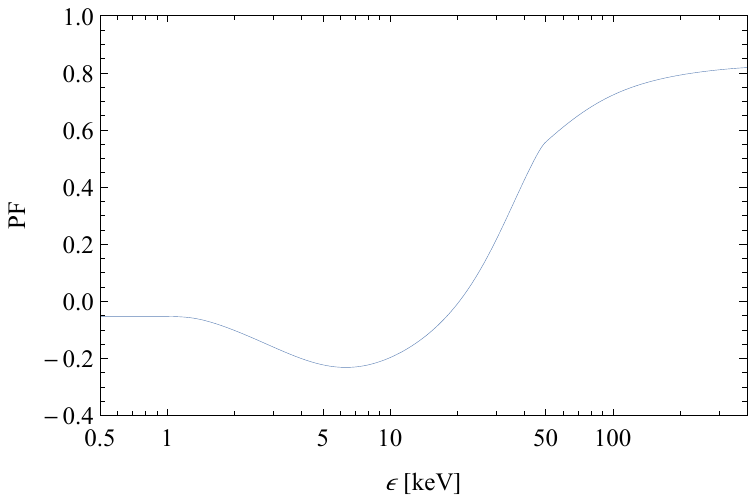}
	\caption{Same as Fig. \ref{fig:str1}, but for \mbox{$v(r_{\rm up})=1.3\times 10^{10}$~cm~s$^{-1}$} (panels for the local spectra are not shown). 		
  		}
  \label{fig:str1v1.3}
	\end{center}
 \end{figure*}

\begin{figure*}
 	\begin{center}
     \includegraphics[width=0.19\textwidth]{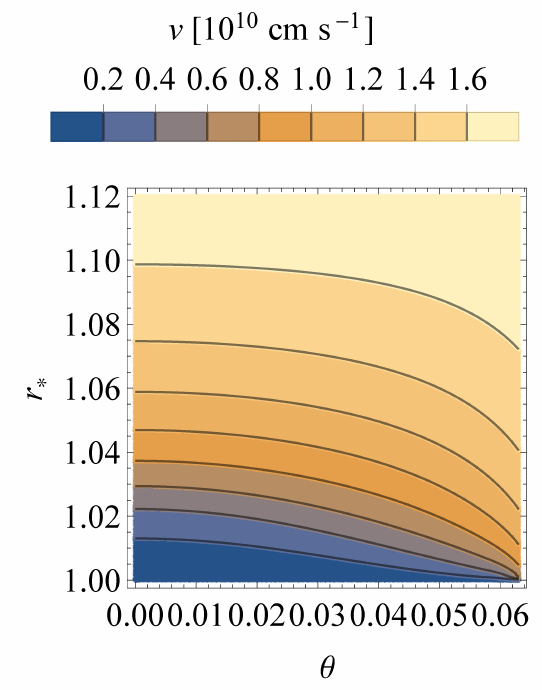}
  \hfill
  \includegraphics[width=0.19\textwidth]{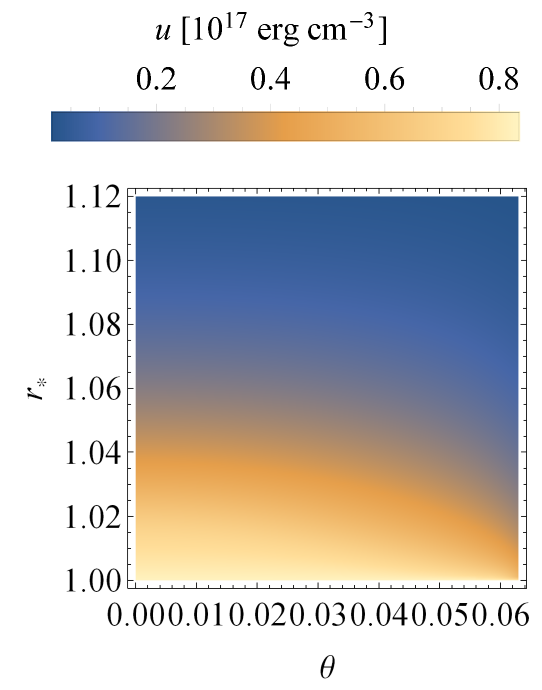}
    \hfill
    \includegraphics[width=0.19\textwidth]{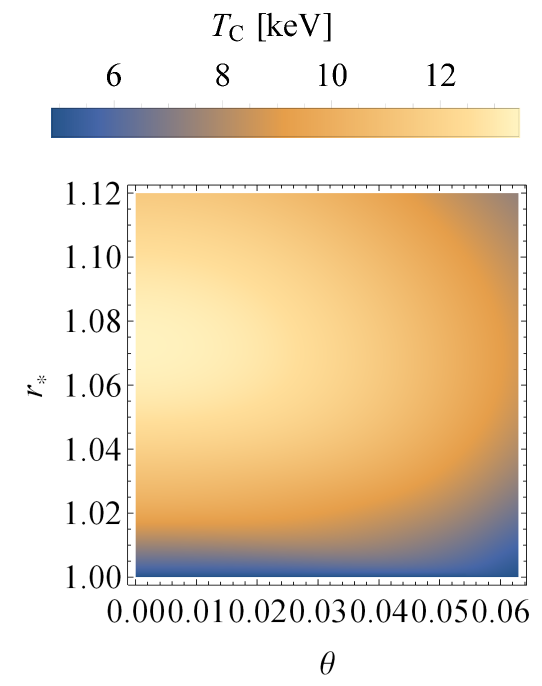}
        \hfill
   \includegraphics[width=0.19\textwidth]{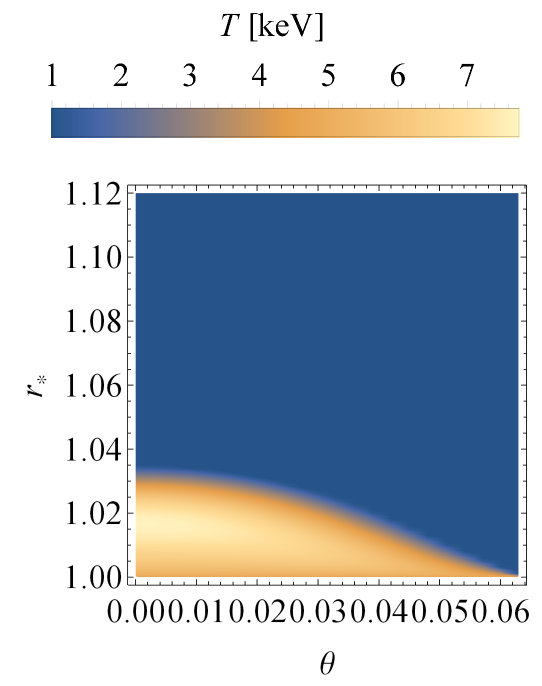}
          \hfill
   \includegraphics[width=0.19\textwidth]{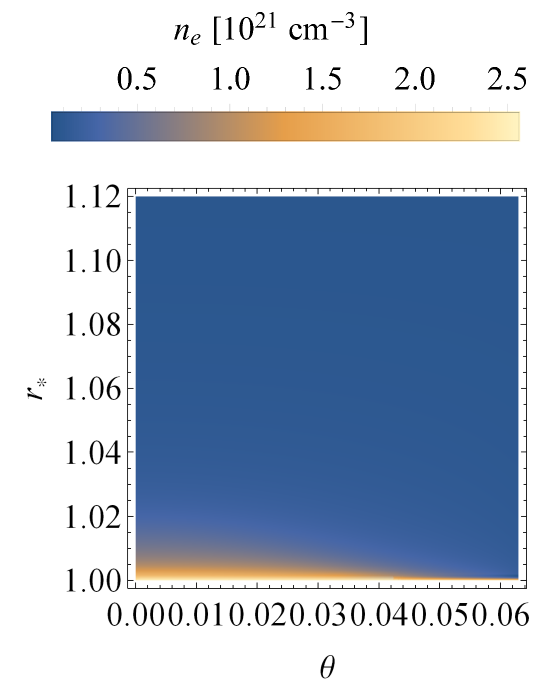}\\\vspace{2mm}
   \includegraphics[width=0.48\textwidth]{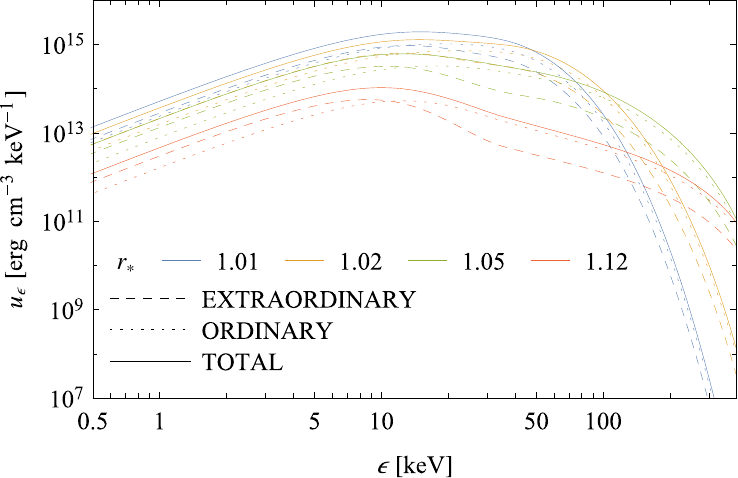}
  \hfill
  \includegraphics[width=0.48\textwidth]{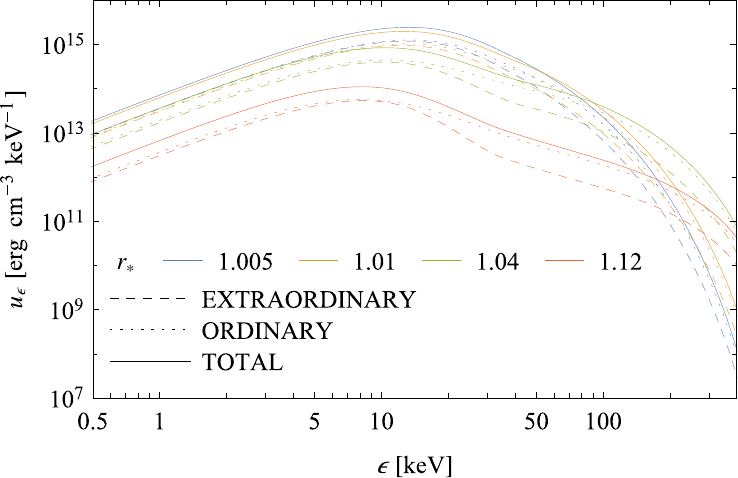}
  \\\vspace{2mm}
   \includegraphics[width=0.48\textwidth]{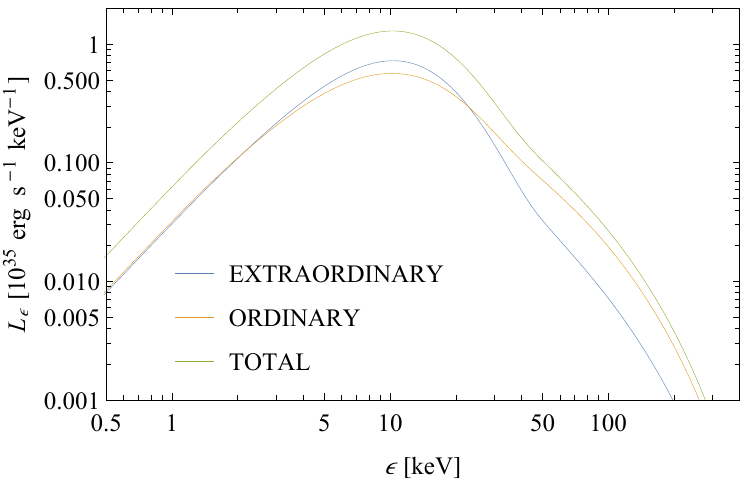}
  \hfill
  \includegraphics[width=0.48\textwidth]{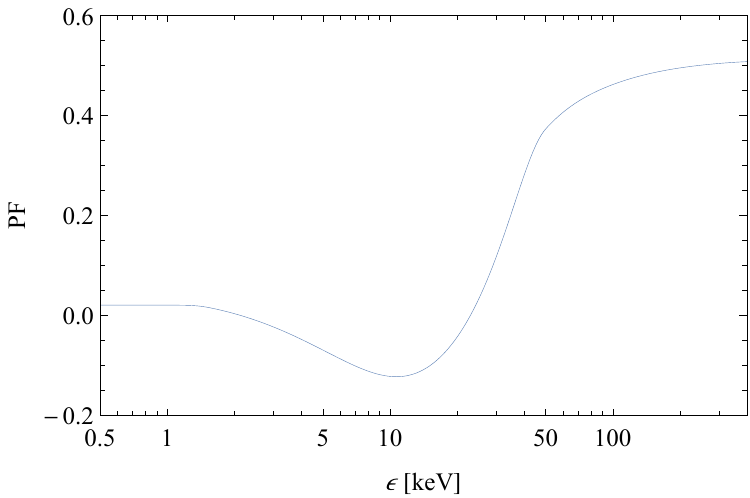}
  		\caption{Same as Fig. \ref{fig:str1}, but for $\dot M_{17}=0.2$. In the right panel, the spectral radiation energy density is shown for \mbox{$\theta=0.054$}.
  		}
  \label{fig:str.2}
	\end{center}
 \end{figure*}

It follows from the description above  that, in simulations, the absolute values of PF increase with increasing accretion rate \mbox{(Figs~\ref{fig:str1}--\ref{fig:str5})}.   That is,  unsaturated Comptonization of the radiation of the ordinary mode can result in an emergent spectrum  comparably close to the spectrum of the extraordinary mode (at least at \mbox{$\dot M_{17}\lesssim 1$}, Figs~\ref{fig:str1}, \ref{fig:strh}, \ref{fig:str1v1.3} \mbox{and \ref{fig:str.2}}).  It is especially clear in the case of the near-critical value of the accretion rate \mbox{$\dot M_{17}=0.2$} \mbox{($r_*(r_{\rm up})=1.15$}, \mbox{$L_{\rm X,\,37}\approx 0.7$)}, see \mbox{Fig.~\ref{fig:str.2}}.  
 The calculated PF  in the range \mbox{$1$--$20$~keV} is in a satisfactory agreement with theoretical that of \cite{1991ApJ...367..575B}  and with characteristic observational values (e.g.~\citealt{2022ApJ...941L..14T,2023ApJ...950...76L}).

 The first vacuum frequency \citep*{1979ApJ...233L.125V} falls in the soft \mbox{X-ray} range in the considered cases. Hence the potential effect of the  conversion of the cross sections between the vacuum frequencies is trivial and reduces to recalling the spectra of the modes above the first vacuum frequency.

   \begin{figure*}
 	\begin{center}
     \includegraphics[width=0.19\textwidth]{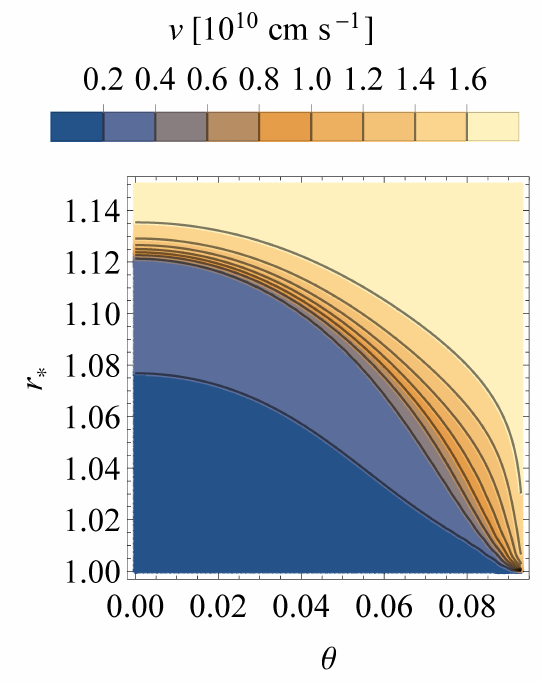}
  \hfill
  \includegraphics[width=0.19\textwidth]{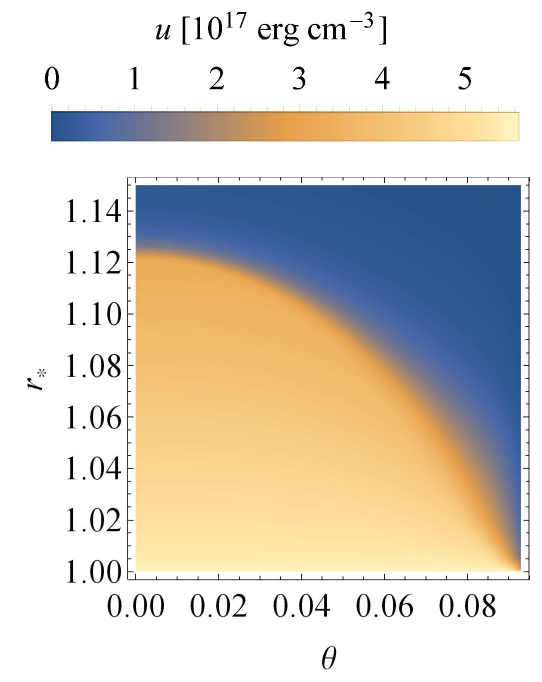}
    \hfill
    \includegraphics[width=0.19\textwidth]{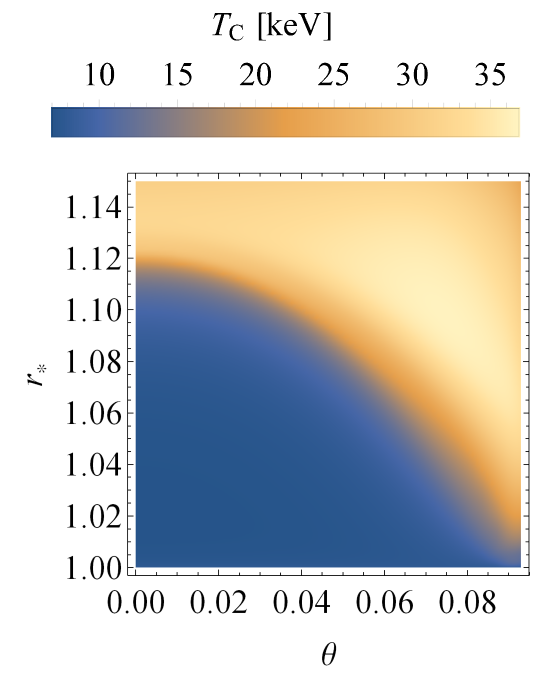}
        \hfill
   \includegraphics[width=0.19\textwidth]{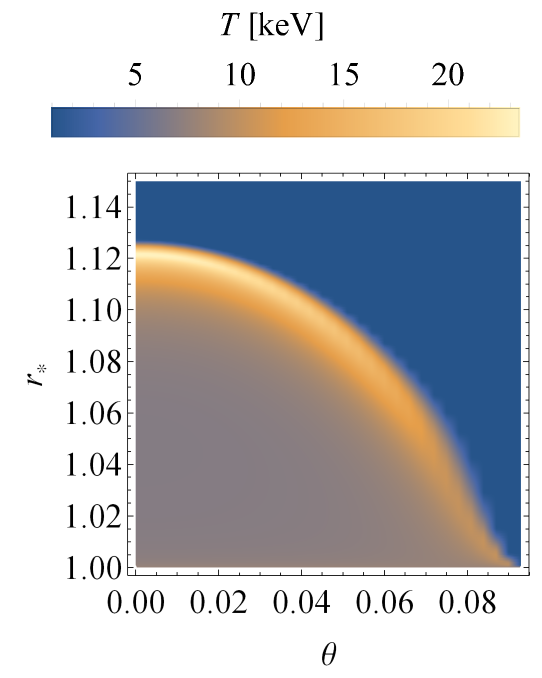}
           \hfill
   \includegraphics[width=0.19\textwidth]{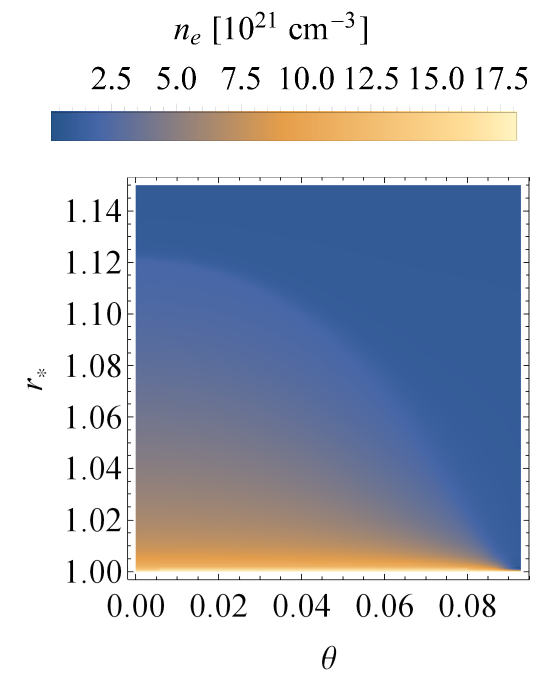}
   \\\vspace{2mm}
   \includegraphics[width=0.48\textwidth]{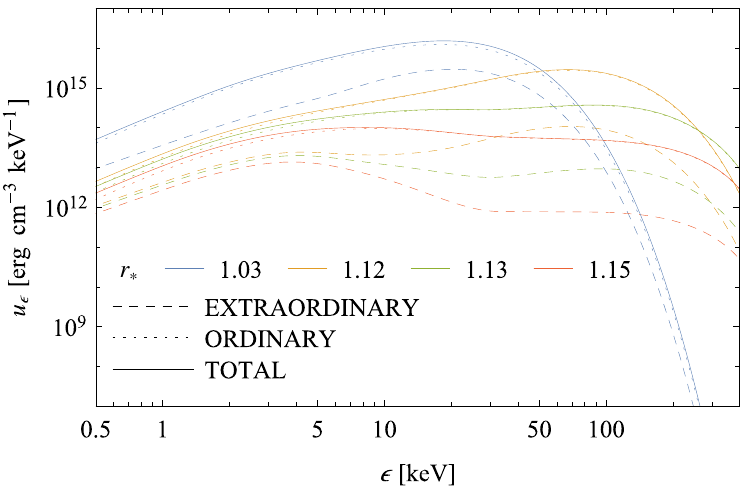}
  \hfill
  \includegraphics[width=0.48\textwidth]{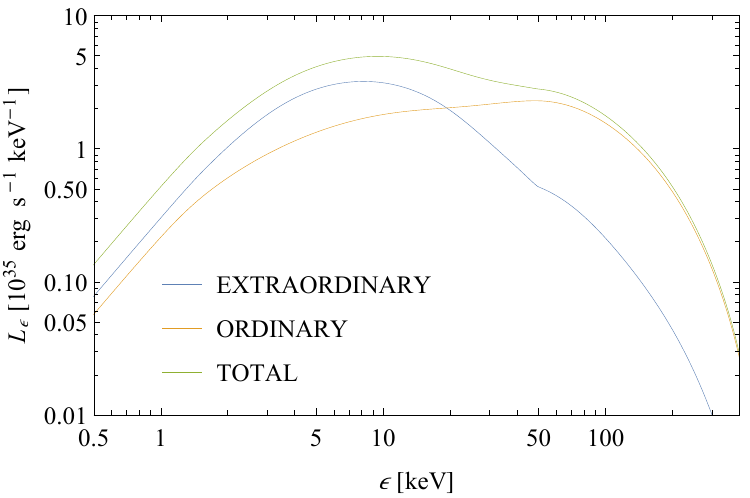}
  		\caption{Same as Fig.~\ref{fig:str3}, but free-free and cyclotron processes are neglected (panels for the local spectra at \mbox{$\theta\neq 0$} and the polarization fraction are not shown).		
  		}
  \label{fig:str3_no_ffj}
	\end{center}
 \end{figure*}

   \begin{figure*}
 	\begin{center}
  \includegraphics[width=0.47\textwidth]{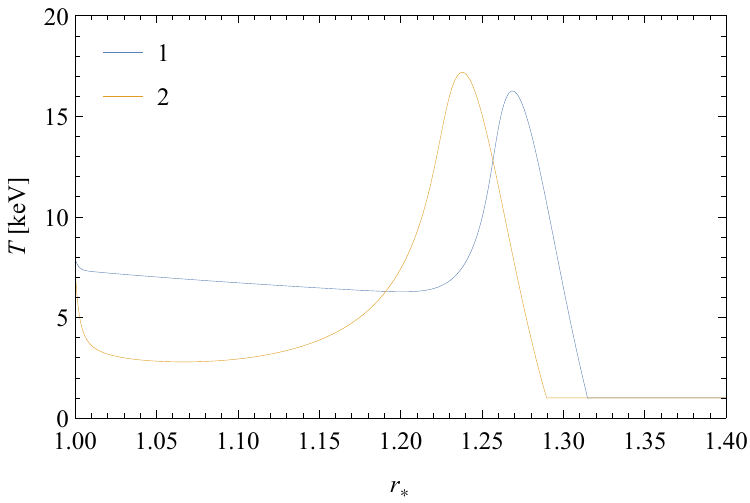}
    \hfill
  \includegraphics[width=0.48\textwidth]{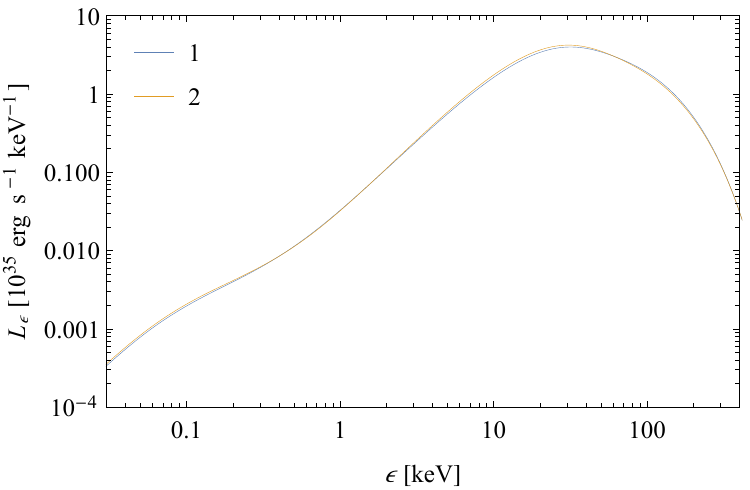}
  		\caption{(Left) the electron temperature  within the filled funnel plotted along the magnetic axis and (right) the emergent spectrum.  The results obtained in the grey approximation for \mbox{$\dot M_{17}=3$}:  \mbox{(1) all} processes are included; \mbox{(2) the induced} Compton effect is ignored. }
  \label{fig:T_r}
	\end{center}
 \end{figure*}

    \begin{figure}
 	\begin{center}
  \includegraphics[width=0.48\textwidth]{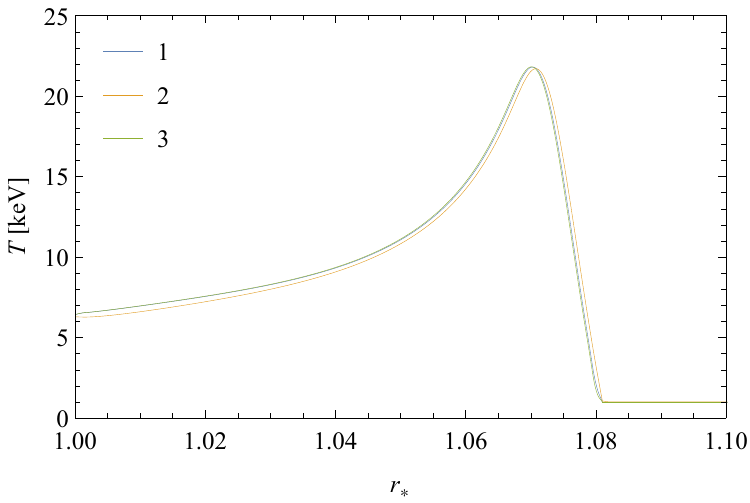}
  		\caption{Electron temperature  within the filled funnel plotted along the magnetic axis. The results of polarization- and spectrum-dependent calculations for \mbox{$\dot M_{17}=1$}: \mbox{(1) all} processes are included (Fig.~\ref{fig:str1}); \mbox{(2) free-free} processes, the cyclotron emission and the induced Compton effect are ignored; \mbox{(3) free-free} processes and the cyclotron emission are ignored.}
  \label{fig:T_rp}
	\end{center}
 \end{figure}

\subsection{Effects of induced scattering and of the magnetic field geometry}
\label{sec:n2} 

 The present simulations  self-consistently show that in general the influence of the induced Compton effect on the  structure of the  accretion column and the radiation spectrum in the funnel cannot be neglected. Inside the settling mound, far enough from the shock wave, Compton scatterings establish a state close to LTE, with induced scattering can play a notable role.  That is, in the specified region  one has
\beq{e:tsettl}
T\approx T_{\rm C}\sim \left(\frac{u}{a_{\rm r}}\right)^\frac{1}{4}\sim T_0,
\eeq
 as it could be expected based on the smallness of the radiation flux there.
The total local spectrum can be rather close to the Planck spectrum or have a characteristic saturated form in this region, but  approach neither the Planck nor the Wien distribution. In the latter case, the spectrum has a zone of less steep slope than in the Rayleigh-Jeans range   \mbox{(e.g. Figs~\ref{fig:str3}, \ref{fig:str5} and \ref{fig:str3_no_ffj})}.

  Fig. \ref{fig:T_r} shows the results for the electron temperature in the filled funnel  along the magnetic axis and for the corresponding emergent spectrum  (cf.~\citealt{1974JETP...38..643Z,1996ApJ...457L..85K}), obtained in a geometrically  2D domain under the grey approximation analogously to the calculations of G2021 (\mbox{section~3}, the system of equations includes the only  transfer equation), but with taking into account the induced Compton effect, second-order bulk Comptonization, free-free processes with the use of \eqn{e:Kff}, the cyclotron emission with the use of \eqn{e:jcyc}, and the dipole geometry of the magnetic field.  The scattering cross section along the magnetic field equal to \mbox{$0.1\sigma_{\rm T}$} is used. 
Such calculations being performed without taking into account the induced Compton effect do not show an agreement of the dependencies for the model Compton temperature in the settling mound \mbox{(Fig.~\ref{fig:T_r})} with the Stefan-Boltzmann law.     
Neglecting  free-free processes leads to an additional decrease in the model Compton temperature. This is in accordance with its values \mbox{$\sim 0.2$--$2$~keV} (at the typical radiation energy density of about \mbox{$2.6\times 10^{17}$--$4\times 10^{17}$~erg~cm$^{-3}$}, which is nearly constant throughout the settling zone)
 appeared in the results of G2021 (\mbox{figs~5} \mbox{and~6}), which thus contain the spectra of a shock wave itself with a soft backlight (cf.~\citealt{1982SvAL....8..330Leng}).

 The grey approximation allows thus modelling  a local situation in the column  that would be rather close to a situation without a strong magnetic field  at least accurate to the details of the shock profiles. It is justified regardless of the value of the scattering cross section along the magnetic field (as long as that value is not too small), the value of the cyclotron emission term and the values of other parameters.
 That is, in a sufficiently weak magnetic field, induced scattering is of great importance for establishing of the electron temperature in the interiors of the settling mound.
 In this case, the value of $T$ is very close to the LTE values throughout the whole zone behind the temperature shock.

It turns out that thermal Comptonization cannot provide the existence of a satisfactory self-consistent steady-state solutions   while the induced Compton effect is ignored,   except for some special cases corresponding usually to relatively low accretion rates \mbox{$\dot M_{17}\lesssim 2$}. 
The decrease in the angle-averaged scattering cross sections with frequency in both modes causes that  in a sufficiently strong magnetic field,  in the models without allowance for induced scattering  (when a solution exists), the model  photons lose the energy in the settling mound slowly, therefore the deficit in the electron temperature  relative to the exact solutions is not dramatic for those models 
and can approach  no much than \mbox{$\sim 1$~keV}  even for the solutions not accounting for the free-free 
processes~\mbox{(Fig.~\ref{fig:T_rp})}.

The value of  $u$ in the settling zone decreases with increasing $r$ the slower, the lesser the  $l$ quantity.  
 The    mound is lower than in the cylindrical channel at the same parameters (the area of the transverse cross section of the cylindrical channel is assumed to be  equal to the area of the funnel base). The radiation flux component along the magnetic field  differs from zero  in the interiors of the  settling mound, consequently, mostly due to the geometry of the magnetic field.

Within the shock wave, the effect of induced scattering is insignificant at any magnetic field, since the $n_i$ quantities are small at the large radiation fluxes. 
The influence of the anisotropy of the photon occupation numbers, which is mostly determined by the squares of those and by the coordinate,  should not then change the results for the electron temperature strongly  (\citealt{1970ApL.....7...69L}).

Taking into account the induced Compton effect in the radiative transfer treatments was performed earlier in the framework of static models of emitting regions \citep*{1989ApJ...337..426M,1989ApJ...342..928A}.

\subsection{Effects of bulk Comptonization}
\label{sec:v2}

The thermal structure of the shock differs from the settling region mainly because of the influence of first-order bulk Comptonization (compressional `acceleration' of photons) \citep{1981MNRAS.194.1033B,1981MNRAS.194.1041B}.
This is one of the mechanisms in addition to the thermal Doppler effect, which supplies photons at high energies and causes the heating of the plasma due to the recoil in the region of great values of \mbox{$|\nabla\cdot\bm v|$} to electron temperatures  considerably exceeding  the values of the  \mbox{$(u/a_{\rm r})^{1/4}$} quantity  and the value of~$T_0$.

Second-order bulk Comptonization occurs mostly in the upper, relatively cold layers of the shock wave and in a rather narrow low-intensity layer of the pre-shock region from where the radiation is still capable of contributing to the emergent flux.   In principle, it is possible to model approximately the structure of the shock by solving self-consistently the transfer equations of sort of equations \eqn{e:transf}, ignoring both thermal and second-order bulk Comptonization --  in such solutions  the emergent spectra of both modes  have the  power-law tails, which is in accordance with the results of \cite{1981MNRAS.194.1041B} and \cite{2005ApJ...630..465B}.

Compared  to the results where second-order bulk Comptonization  is neglected, the peak height of the calculated shock above the neutron star surface is relatively larger  (the typical difference is described by a factor of~$2$), while the calculated electron temperature in the shock wave is relatively higher  (and typically differs  by \mbox{$1.5$--$2$} times), which indicates the increase in the density of  hard photons (forming eventually the tail of the spectrum of the funnel). 
 The quantities $T$ and \mbox{$T+{m_{\rm e}v^2}/{(3k)}$} become almost equal at the value of $r$ (at a given~$\theta$) at which the value of $T$ approaches its maximum.   The corresponding contour is close to the inner boundary of the shock wave.   Above that contour, one can distinguish another level where the dominance of the \mbox{$m_{\rm e}v^2/3$} term   is relieved by the dominance of the electron temperature.
 An excess of hard photons in the upper part of the settling mound manifests in a spreading of the inner boundary of the temperature shock or (in relatively weak fields)  in a smaller altitude of this boundary in comparison to the velocity shock.  The $T_{\rm C}$ quantity approaches a maximum in the outer layers of the shock, which indicates that as the hard radiation penetrates against the accretion flow into the region above the shock wave, losing in the photon number density.

 In the absence of a magnetic field, first-order bulk Comptonization would be the reason of  a shift in the maximum of the emergent spectrum towards higher energies relative to the characteristic energy~\mbox{$2.8kT_0$}. This is a typical picture (see \citealt{1999ASPC..161..410P}, \mbox{fig.~1}).    In a strong magnetic field this effect takes place as well.
  The emergent spectrum of the extraordinary mode is shifted to higher energies relative to the value of \mbox{$kT_0$} mainly due to the influence of first-order bulk Comptonization, while the thermal effect of the shock wave is not very significant for the shift. The spectrum of the ordinary mode can be affected by the heating of the shock wave.  If a high-energy hump in the emergent spectrum of the ordinary mode is formed, it usually corresponds to the energy~\mbox{$\sim 2kT_{\rm s}$}, where $T_{\rm s}$ is the maximum electron temperature (approaching in the shock wave).

 Thus, second-order bulk Comptonization leads to an increase in the intensity in the high-energy tails (both directly and through the heating of the medium) compared to the tails formed by first-order bulk and thermal Comptonization.
 The high-energy tail of the total spectrum of intensities well above those of the extraordinary mode is formed as well in the framework of self-consistent calculations with an artificially cooled shock wave to a temperature of $T_0$ or to  a temperature taken to be equal \mbox{to~$(u/a_{\rm r})^{1/4}$}. The high column in such simulations is not formed, so that the resulting `shock wave' lies on the neutron star surface~-- this shows  that the heating of the shock wave at least in the correspondence with solutions of G2021 is the necessary condition for the  formation of the dynamical structure (in contrast to the heating of the post-shock zone).

The influence of the \mbox{$m_{\rm e}v^2/3$}  term  can be illustrated by varying the velocity at the upper boundary. Consider a filled funnel. In \mbox{Fig.~\ref{fig:str1v1.3}} the structure, the emergent spectra and the polarization fraction are shown for \mbox{$v(r_{\rm up})=1.3\times 10^{10}$~cm~s$^{-1}$} \mbox{($\dot M_{17}=1$}, \mbox{$L_{\rm X,\,37}\approx 1.7$)}.  The decrease in \mbox{$v(r_{\rm up})$} leads to the shift of the high-energy cutoff towards low energies, as expected.
   A considerable decrease in \mbox{$v(r_{\rm up})$} (up to \mbox{$\sim c/3$})  demands a decrease in $\gamma$ for the existence of solutions in the framework of the model. Although such solutions contain a significantly cooler shock (up to \mbox{$T_{\rm s}\approx 9$ keV} at \mbox{$T_0\approx 7$ keV} for \mbox{$\dot M_{17}=3$}), the relatively softer spectral tail remains nevertheless rather hard, and the PF distribution is similar to that given in Fig.~\ref{fig:str3}.  
 The possibility of the distinction of \mbox{$v(r_{\rm up})$} from the free-fall velocity was mentioned by \cite{2013ApJ...777..115P}.
 
 Within the models constructed under the grey approximation, the influence of the  \mbox{$m_{\rm e}v^2/3$} term  on the formation of the emergent spectrum (see an example in \mbox{Fig.~\ref{fig:T_r}}) manifests by adding a hard-energy tail  to the typical quasi-Bose-Einstein distribution.

\subsection{Free-free processes and the cyclotron emission}
\label{sec:jff}

 Free-free processes  cause the increase in the brightness temperature of the emergent spectrum at the energies  less  \mbox{$\sim 0.5$~keV}  in comparison to the case of pure scattering.  It is shown clearly in the grey approximation \mbox{(Fig.~\ref{fig:T_r})}, in the framework of which the low energies are considered.
 The density of matter in the post-shock region is insufficient to make the free-free absorption an efficient mechanism of  heating the plasma. Figs~\ref{fig:str3_no_ffj} and~\ref{fig:T_rp} represent the results of computations with free-free and cyclotron processes neglected. It is clear from these results that neglecting free-free processes within the model does not vary the output distributions strongly and, thus,     that the Compton energy exchange under the conditions of the local Compton equilibrium is the main mechanism of the heating of the plasma in the column~\mbox{(\S~\ref{sec:n2})}.

  A more pronounced manifestation of free-free interactions could be obtained with the parameters differ from used here. This will be studied in the separate work.  The $j_{{\rm cyc,}\,i}$ terms cause only 
a small relative increase in the height of the mound \mbox{(up to \mbox{$\sim 5$\%})} compared to the solution for the terms neglected. The combined action of free-free processes and the cyclotron emission leads to the model structures of a practically identical height  for the cases considered (Figs~\ref{fig:str3} and~\ref{fig:str3_no_ffj}), which is not the case in general (although there are no major differences expected).

 Notice that the contribution of the overall seed radiation to the radiation energy density at the bottom is included 
 in~\eqn{e:bottemp}.
Therefore, $\gamma$  should insignificantly differ for the case of neglecting free-free processes inside the column \mbox{(Fig.~\ref{fig:str3_no_ffj})}. The corresponding difference is small and neglected.

 \section{Discussion} \label{sec:d}
 
 \subsection{The model}

The overall picture of matter breaking in the radiation-dominated regime is thus as follows. Five zones can be  distinguished in the accretion channel above the magnetic pole: \mbox{(1) the cold} accretion flow falls near-freely in a radiation field of relatively low energy densities; \mbox{(2) the flow} begins to fluently brake in the shock, but its electron temperature is not yet very high; predominantly, the effect of second-order bulk Comptonization in this zone leads to the formation of the tails in the emergent continuum; \mbox{(3) the flow} experiences the rather sharp braking in high-density layers of the shock wave up to a near-complete stop,  photons gain the energy in this zone due to first-order and second-order bulk Comptonization, and by thermal Comptonization; the plasma can be heated  to \mbox{$10$--$20$~keV} and higher by means of the bulk-heating mechanism there;  \mbox{(4) the matter} falls, slowing down further, within  a settling mound surrounded by the shock wave and the neutron star surface; as the matter settles, the electron temperature gradually approaches thermodynamically equilibrium one  and becomes typically \mbox{$1.5$--$4$} times less than the temperature in the shock wave; \mbox{(5) the plasma} penetrates through the magnetic field beyond the channel near the base of the emitting region.

The assumption of LTE is a typical example of that which cannot actually work in the shock wave.  Thus, \cite{2021MNRAS.508..617Z} have used the Stefan-Boltzmann law to write the radiation temperature for the balance equation, which caused the obtaining of too low electron temperature in the shock, where the Compton heating and cooling must be modelled, with  the bulk-heating mechanism \citep{1981MNRAS.194.1041B} must be taken into account. 
 The velocity profiles in the solutions without allowing for modelling thermal Comptonization are not very sensitive  to the assumption on the electron temperature, since the radiative diffusive and enthalpy fluxes must not be strongly dependent on it \citep{1973NPhS..246....1D, 1987ApJ...312..666A}. Corresponding transfer equation for the total radiation energy density is obtained from the transfer equation for the spectral radiation energy density under the assumption about the Compton equilibrium (Appendix~\ref{app:eur}).

The system of equations analysed by \cite{2021MNRAS.508..617Z} (appendix~A) corresponds to equations of G2021, but accounts for the effects of the gravitational acceleration and viscosity (the scattering is assumed to be isotropic). The  viscosity terms are neglected during the consideration of the origination of photon bubbles in the WKB approximation.
Since the development of photon bubble instability is related to perturbations of the mass density and total radiation flux, its description should not be strongly dependent on the particular results for the electron temperature.

The process of the going out of the plasma beyond the column needs to be modelled in detail in the framework of MHD simulations. 
Simple estimates were made by  \cite{1988SvAL...14..390L}. One can obtain that the mass growing above the neutron star crust can originate between outpourings a dense gas-dominated non-steady thermalized layer of the height \mbox{$Z\lesssim 2kT{R^2}/({GMm_{\rm p}})\approx 60 ~{\rm cm}$} (for the constant temperature taken to be equal \mbox{$6$~keV}).
This layer, or `the thermal mound' \citep{1998ApJ...498..790B}, can be ignored in the calculations of the radiation transfer in the funnel, since its altitude is small and
  the  energy exchange throughout the medium above the layer is significant. 


The radiation energy density corresponding to the post-shock velocity equal to \mbox{$1/7$} of the pre-shock value cannot serve for the implementation of the bottom boundary condition, since this post-shock value of the velocity was obtained for a typical infinite shock wave (e.g.~\citealt{1967pswh.book.....Z}). 
The circumstances of \mbox{X-ray} pulsars (e.g.~the boundedness of the settling mound by the neutron star surface, the  transverse boundedness of the gas flow, the presence of a magnetic field)  lead to \mbox{$\gamma\sim 1$}.
 The used value of $\gamma$  leads to  \mbox{$v(R)$} of about \mbox{$3\times 10^{8}$--$4\times 10^{8}~{\rm cm~s^{-1}}$}.  A reasonable variation in $\gamma$ in both directions cannot change the results considerably (for 
  \mbox{$\gamma>0.99$} a steady solution does not often exist~yet).

The model described in Sections~\ref{sec:calc} and \ref{sec:res} does not imply the consideration and solution of the energy equation for the total radiation energy density (Appendix~\ref{app:eur}). One can guess that its inclusion and thus the transition to solving two separate problems (obtaining the velocity profile, e.g.~\cite{1973NPhS..246....1D}, and then  self-consistent calculation of the electron temperature and spectral radiation energy density) may appear  more useful in some studies. 
 Nevertheless, the method described in the present paper guarantees full compliance of all obtained  distributions in the framework of hydrodynamical approach,  
 providing better, as well as joint, fulfilment of the \mbox{Liouville} theorem.  The self-consistent solutions take into account the result of calculating the realistic frequency photon redistribution, in contrast to the spectrum-integral  approaches including the equations of a form of those of Appendix~\ref{app:eur}.

 As is clear from above, in the calculations  aimed to obtain the dynamical structure and  carried out with the use of the grey approximation  it would be reasonable to use the scattering cross section along the field pretty close to the Planck  and Rosseland means of the $\sigma_1$ quantity  to achieve correct characteristic values of~\mbox{${\nabla\cdot\bm v}$}.  The Planck mean reads \mbox{$\langle\sigma_1\rangle_{\rm P}=40/21\,({\pi kT}/{\epsilon_{\rm cyc}})^2\sigma_{\rm T}$} for \mbox{$\epsilon_{\rm cyc}/(kT)$} well exceeding \mbox{$\pi\sqrt{2}$}, and the Rosseland mean reads
  \mbox{$\langle\sigma_1\rangle_{\rm R}=4/5\,(\pi kT/\epsilon_{\rm cyc})^2\sigma_{\rm T}$} for
   \mbox{$\epsilon_{\rm cyc}/(kT)$} well exceeding~$\pi$. The  polarization-averaged Planck and Rosseland means calculated for the accurate magnetic opacities were presented by \cite{2022MNRAS.517.4022S}.

  To obtain the results shown in Fig.~\ref{fig:T_r}, the less scattering cross section along the field  compared with the specified means was used for computational reasons, which caused relatively more gentle profiles of the velocity within the shock.
  The use of small values \mbox{$\sim 10^{-2}\sigma_{\rm T}$} for that cross section   would lead to a   very smooth shock profile, so that the settling mound would be virtually absent. This means that there is no the same grey equivalent for obtaining  simultaneously for the case of a strong magnetic field both the structure and the total spectrum, whose slope of the quasi-exponentially fallen power-law  corresponds at characteristic values of the total \mbox{$y$-parameter} to a grey cross section along the field much smaller than~\mbox{$10^{-2}\sigma_{\rm T}$}.

The main drawback of the present work is the lack of calculations of the radiative transfer at frequencies of the cyclotron resonances, whose influence remains to be  unstudied. In general, those calculations should be associated with the accurate scattering and absorption cross sections for a hot magnetized plasma calculated in the temperature range characteristic for the column 
(e.g.~\citealt{2017A&A...597A...3S, 2022MNRAS.517.4022S, 2022PhRvD.105j3027M, 1992hrfm.book.....M}).  In principle, it is not difficult to incorporate into the codes the cross sections containing the cyclotron resonance (resonances) instead of the used  cross sections, for which there is the only nuisance associated with the corresponding glaring retardation of the convergence. 
 Another lack of the model developed is the description of the radiative transfer in the non-relativistic Fokker-Planck approximation. Comptonization is modelled using the angle-averaged cross sections, which is rather crude applied to the ordinary mode (variation of this quantity in the pretty wide range, however, does not affect the results dramatically). Solving a system of equations similar to that described in Section~\ref{sec:calc} and including the integro-differential transfer equations written for a moving medium of the funnel  instead of expansions \eqn{e:transf} would yield  more accurate results containing  angular distributions of the radiation intensity. Furthermore, the relativistic description 
of the radiative transfer should be used.

Although the photon energy dependence of the cross sections for both modes plays a crucial role in the spectrum formation, the role of terms containing the photon energy derivatives of the angle-averaged cross sections is insignificant.
The mode  switching is taken into account using the results of studies on the coupled-diffusion approximation of the radiative transfer \citep*{1981ApJ...251..278N, 1982Ap&SS..86..249K, 1992hrfm.book.....M}.

A few remarks should be made on the alternative calculations. A drawback of the  existing 1D numerical and analytical models is that, in the framework of them, a significant part of the photons leave the column directly from the sidewall of the zone of the slowly falling gas, without crossing the shock wave, as was noted above. This process is often described  by a term in the transfer equation written by means of the averaged diffusion time across the column \citep{2007ApJ...654..435B, 2016A&A...591A..29F,2017ApJ...835..130W}. A similar simple picture is  typical for  dynamical models of \cite{1976MNRAS.175..395B} and \cite{1977A&A....60...39T} constructed by means of averaging the radiation flux across the magnetic field. 
 Calculations of the emergent spectrum  with the use of 1D models may take into account the dynamical effects in the insufficient extent.   Such models open a way to fit easily a vast amount of observations.

The solutions of \cite{1986Ap.....25..577L}   were obtained under the assumption of a fenomenological dependence of the photon occupation number on the optical depth. The shape of the obtained spectra at frequencies less than the frequency of maximum is artificial. The dynamical effects were  not considered. These results   led to the lack in important details in describing the formation of a continuum carried out by \cite{2015MNRAS.452.1601P} (the radiation of the ordinary mode advected downwards in the shock wave was totally ignored).

\subsection{On the fit of observations}

The high-energy tails, expanding mainly due to the influence of  second-order bulk Comptonization, can become an obstacle to fit  observed fallen quasi-power-law spectra with the developed model.    The received flux in the tails is often relatively low (e.g.~\citealt{2017A&A...607A..88K}).
If future simulations taken into account angular photon redistribution and cyclotron resonances do not lead to better agreement with observations at usual problem parameters, this will mean that either the velocity at the top boundary may be distinguished noticeably from the free-fall velocity (and then the simplest solution is in the parametrization of the velocity at the top) or the realistic profiles of the shock are much more complicated than obtained here and must be found in the  spectral radiative transfer simulations coupled with the MHD simulations.  An experimental explanation may consist in the absence of the corresponding observational data.   The obtained hardness of the spectrum  taken near the energies $10$~keV and $1$~keV may exceed the observational value because of ignoring the reproduction of the radiation in the atmosphere of a neutron star, the albedo of which is characterised by a low-frequency cutoff \citep{2015MNRAS.452.1601P}.

Recently, the observed continuum of the radiation from the \mbox{X-ray} pulsar \mbox{1A 0535+262}   generated during the giant outburst and   having a high-energy spectral maximum at energies of \mbox{$\sim 20$--$30$~keV} has been described by  \cite{2022ApJ...932..106K} and \cite{2024MNRAS.528.7320S}. Another well-known source with similar spectral properties is \mbox{V 0332+53} \citep{2017MNRAS.466.2143D}.
As an interpretation, one can consider the radiation with a spectral maximum in the vicinity of the maximum of the spectrum of the extraordinary mode from a strongly heated emitting region of a relatively small area (bulk Comptonization can contribute in the displacement of the maximum as described above). The approximate examples of the continuum of the corresponding form are shown in \mbox{Fig.~\ref{fig:strh}}.    The significant negative correlation of the CRSF energy with the \mbox{X-ray} luminosity within such a hot spectrum, being considered in the model directly relating the CRSF centroid energy to the height of the column,  would imply a significant absolute difference in the height of the column in the considered states, which cannot be demonstrated by the narrow hollow column. (The relatively small height of the hollow columns was previously obtained in numerical simulations of \citealt{2015MNRAS.452.1601P}.)  The possible solution seems in taking into consideration the distortions in the magnetic field strength occurring in the column (see examples of solutions of the Grad-Shafranov equation for preset mounds obtained by \citealt{2013MNRAS.430.1976M} and \citealt*{2013MNRAS.435..718M}). Weaker magnetic fields appearing with increasing $\dot M$ would explain the observed anticorrelation of the CRSF energy with the \mbox{X-ray} luminosity. The check is to become the aim of further self-consistent MHD simulations.

Preliminary simulations in the current model with the surface value of the magnetic field strength less than considered above (it was set \mbox{$\epsilon_{\rm cyc}(R)=10$~keV})  lead to the two-hump total emergent spectrum with the location of the CRSF in the spectral saddle. The exact spectral shape is to be modelled in the framework of upcoming studies, which will also include the effects of  complicated (non-dipole) geometry of a magnetic field.

Theoretical models of the column with a collisionless shock wave also lead to solutions corresponding very strong heating of the plasma (often up to tens of~keV) during passing the shock wave \citep{1982ApJ...257..733L,2004AstL...30..309B}. In these models, the  hot material fills the entire post-shock zone up to the neutron star atmosphere, which may be another problem for the interpretation of the observational data. In the calculations of \cite{2017A&A...601A.126V} using the Feautrier method, the electron temperature was parametrized  to fit the observational spectra and appeared to be rather low \mbox{($5$--$6$~keV)}.  Detailed self-consistent modelling of the spectrum of the radiation from the column with the collisionless shock is the problem that remains to be solved.

The formation at  temperatures \mbox{$\sim T_{\rm eff}$--$T_0$} and under  circumstances not considered here of the spectra having the maximum in corresponding to these temperatures spectral range and a  high-energy power-law  should be studied further.

\section{Conclusions} \label{sec:concl}

The accretion regime on to a strongly magnetized neutron star corresponding to luminosities above the value of $L_{\rm cr}$ has been considered. The main processes within the magnetic polar funnels have been investigated.
The transfer equations for the polarized radiation  have been numerically solved within the considered domains self-consistently   with the expressions for the electron temperature and the radiation energy density, the momentum equation, and the continuity equation.  The solutions obtained describe the dynamical and thermal structure both of the radiation-dominated shock wave and the settling mound, and the polarimetric \mbox{X-ray} continuum of radiation that emerges from the funnel.

It has been found that  the peak height of the model accretion column, being calculated taking into account the specific form of the frequency- and polarization-dependent scattering cross sections, can depend nonlinearly  on the accretion rate, in contrast to the nearly linear outputs obtained in the grey approximation. In the case of a sufficiently strong magnetic field,  the saturation in the column peak height--accretion rate dependence takes place in the studied range of the accretion rate.

The significance of the Compton scatterings within the accretion column is as great as it is characteristic, for example, for the early universe. 
 Comptonization has been shown to be the main process for the formation of the temperature structure of the column.
The electrons are in the thermodynamic equilibrium with the radiation deep below the shock wave, as  proofed in the way of using the Compton-equlibrium formula for obtaining the electron temperature, which noticeably exceeds the blackbody value within the shock due to the contribution of the effects of bulk motion to the heating of the plasma.

The temperature structure should be calculated in general case with taking into account the influence of the induced Compton effect, which impact on the electron temperature within the settling mound  under the condition of the sufficiently weak magnetic field and (at the sufficient accretion rate) on the existence of the steady-state solutions of the problem. The effects of second-order bulk Comptonization are  important mainly for the formation of the spectra of emergent radiation, resulting in  a smoothed power-law high-energy zone with a cutoff.  I have discussed  main directions for updating the results in the framework of  more accurate models.

 The calculated  absolute value of the polarization fraction lies between \mbox{$0.05$} and \mbox{$0.5$} being measured near the spectral maximum of the extraordinary mode.
The model developed can provide the emergent continuum of the ordinary mode close to that of the extraordinary mode.
 Namely, the  calculations for the configuration of a geometrically thin hollow column can lead to the spectrum of the funnel characterised by a high-energy maximum and a low polarization fraction over the entire spectral range. The similar picture for the cooler spectra is typical for the filled radiation-dominated column at sufficiently low accretion rates  below \mbox{$\dot M_{17}=1$} that are close to the critical value.

\section*{Acknowledgements}
I thank R. Staubert and R. Rothschild \mbox{for their responses}.
The work was supported by the Foundation for the \mbox{Advancement} of Theoretical Physics and Mathematics \mbox{`BASIS' (grant 17-15-506-1)}.

\section*{Data availability}
The calculations described in this paper where performed using the private codes developed by the author. The data presented in the figures are available on reasonable request.

 \bibliographystyle{mnras}
\bibliography{app3}

\appendix

\section{Equations for the total radiation energy density}\label{app:eur}

The physical model considered above  does not contain any of transfer equations for the total radiation energy density. Nevertheless, let us consider them, limiting to consideration regardless of the polarization of radiation.

 Let free-free and cyclotron processes be neglected and the term of the second order in the bulk velocity be ignored. Then it follows from the transfer equation \citep{1981MNRAS.194.1033B} that  in a steady situation (the subscript `np' for the occupation number and radiation energy density will be dropped)
\beqa{e:transfint}
\nabla\cdot(\hat D\nabla u)-\bm v\cdot\nabla u-\frac{4}{3}u\nabla\cdot\bm v\\\nonumber + \frac{\sigma n_{\rm e}}{m_{\rm e}c}u\left(4kT-\bar\epsilon-\frac{\int\limits_0^\infty\epsilon^4n^2\rm d\epsilon}{\int\limits_0^\infty\epsilon^3n\rm d\epsilon}\right)=0,
\eeqa
where $\hat D$ is the diffusion tensor, the cross section $\sigma$ is represented analogously to expression (25) of \cite{1981MNRAS.194.1033B} with accuracy to accounting for the $n^2$ term, and
\beq{}
\bar\epsilon=\frac{\int\limits_0^\infty\epsilon^4n\rm d\epsilon}{\int\limits_0^\infty\epsilon^3n\rm d\epsilon}.
\eeq
Using \eqn{e:temperZL} one can obtain from \eqn{e:transfint}
\beq{e:en1}
\nabla\cdot\bm F=\frac{1}{3}\bm v\cdot\nabla u,
\eeq
where
\beq{}
\bm F=-\hat D\nabla u+\frac{4}{3}u\bm v
\eeq
is the total radiation flux.
If the momentum equation is considered in the form \eqn{e:mom}, then  from \eqn{e:mom} and  \eqn{e:en1} one can obtain  \citep{1973NPhS..246....1D}
\beq{e:end}
\nabla\cdot\bm F=-\frac{1}{2} n_{\rm e} m_{\rm p}\bm v\cdot\nabla v^2.
\eeq
If the gravitational acceleration is not neglected, the momentum equation reads \citep{1981A&A....93..255W}
\beq{eq:momgr}
n_{\rm e}m_{\rm p}({\bm v} \nabla)\bm v=-\frac{\nabla u}{3}-\frac{GMn_{\rm e}m_{\rm p}}{r^2} \frac{\bf r}{r},
\eeq
where ${\bf r}$ is the vector directed from the neutron star, the length of which is counted as $r$; the difference from the usual distance to the neutron star centre is neglected in the last term for the values of $r$ under consideration (because of the direction of the velocity, where appropriate, one could limit oneself to using the derivatives with respect to $r$ and the  $v$ quantity).
Collecting \eqn{e:en1} and \eqn{eq:momgr} yields \citep{1981A&A....93..255W}
\beq{e:enwf}
\nabla\cdot\bm F=-n_{\rm e} m_{\rm p}\bm v\cdot\nabla\left(\frac{v^2}{2}-\frac{GM}{r}\right),
\eeq
which can easily be written out straightforwardly from the energy conservation law, as well as~\eqn{e:end}.

 These equations are  crudely valid while the flow is not too dense, and hence  free-free processes are not important.  
   Let the second-order bulk Comptonization term be now taken into account.
 Without specification of the terms taken into account in  the right-hand side of the momentum equation and  of the way of the calculation of $T$, one has for the non-steady case (e.g.~\citealt{2017ApJ...835..130W})
\beq{}
\nabla\cdot\bm F+\frac{\partial u}{\partial t}=\frac{1}{3}\bm v\cdot\nabla u+\frac{4\sigma n_{\rm e}u}{m_{\rm e}c}\left(k (T-T_{\rm C,\,np})+\frac{m_{\rm e}v^2}{3}\right)+S,
\eeq
where $t$ is  time and $S$ may include the terms responding for free-free and cyclotron processes.

\label{lastpage}
\end{document}